\documentclass[fleqn,usenatbib]{mnras}

\usepackage[T1]{fontenc}
\usepackage{color}
\usepackage{xspace}
\usepackage{subcaption}

\usepackage{graphicx}	
\usepackage{pgf}	
\usepackage{amsmath}	
\usepackage{amssymb}	
\usepackage{xifthen}
\usepackage{placeins}
\usepackage{textgreek}   
\usepackage{adjustbox}  

\usepackage{hyperref} 

\usepackage{etoolbox}
\makeatletter
\newcommand\sendemail[3]{
\edef\@tempa{mailto:#1?subject=#2 }%
\edef\@tempb{\expandafter\html@spaces\@tempa\@empty}%
\href{\@tempb}{#3}}

\catcode\%=11
\def\html@spaces#1 #2{#1
\catcode\%=14
\makeatother

\newcommand{\todo}[1]{\textcolor{magenta}{[#1]}}
\newcommand{\orcid}[2]{\href{http://orcid.org/#2}{#1}}
\newcommand{\orcidsymb}[2]{\href{http://orcid.org/#2}{#1\adjustbox{trim={-.15\width} {0\height} {-.15\width} {0\height},clip}{\includegraphics[height=10pt]{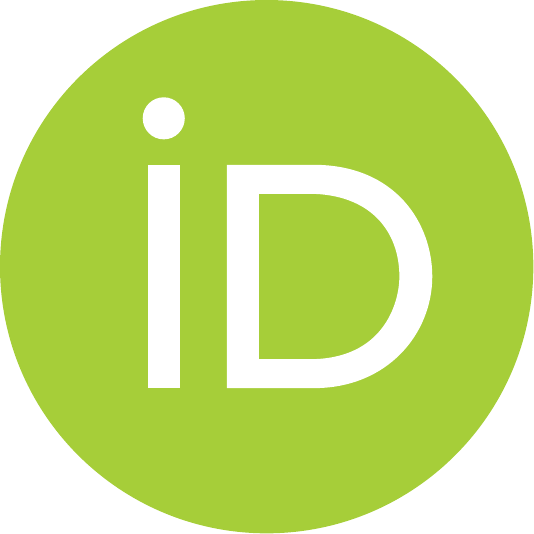}}}}

\newcommand{\sigintr}{$\sigma_\mathrm{intr}$\xspace}
\newcommand{\texp}[1][exp]{\ensuremath{t_\mathrm{#1}}\xspace}
\newcommand{\sigap}{\ensuremath{\sigma_\mathrm{ap}}\xspace}
\newcommand{\spindex}{\ensuremath{I_\mathrm{age}}\xspace}
\newcommand{\sigenv}{\ensuremath{\Sigma_5}\xspace}
\newcommand{\Dr}{\ensuremath{\Delta\,r}\xspace}
\newcommand{\ngroup}{\ensuremath{N_\mathrm{group}}\xspace}
\newcommand{\Halpha}{\text{H\textalpha}\xspace}

\newcommand{\afe}{\text{[\textalpha/Fe]}\xspace}
\def\utd{\ensuremath{u_\mathrm{3d}(\mathrm{dist})}\xspace}
\def\ufd{\ensuremath{u_\mathrm{4d}(\mathrm{dist})}\xspace}

\let\oldtextsigma\textsigma
\renewcommand{\textsigma}{\oldtextsigma\xspace}
\let\oldAA\AA
\renewcommand{\AA}{\text{\oldAA}\xspace}

\defcitealias{deugenio+2021}{DE21}
\defcitealias{magoulas+2012}{M12}
\defcitealias{howlett+2022}{H22}

\title[4-d `fundamental' hyperplane]{The hyperplane of early-type galaxies: using stellar population properties to increase the precision and accuracy of the fundamental plane as a distance indicator}

\author[\orcid{F. D'Eugenio}{0000-0003-2388-8172}~et al.]{\parbox{\textwidth}{
\orcidsymb{Francesco D'Eugenio}{0000-0003-2388-8172}$^{\hyperlink{aff1}{1},\hyperlink{aff2}{2},\hyperlink{aff3}{3}}$\thanks{E-mail: francesco.deugenio@gmail.com},
\orcidsymb{Matthew Colless}{0000-0001-9552-8075}$^{\hyperlink{aff10}{4},\hyperlink{aff5}{5}}$,
\orcidsymb{Arjen van der Wel}{0000-0002-5027-0135}$^{\hyperlink{aff3}{3}}$,
\orcidsymb{Sam P.~Vaughan}{0000-0003-2265-7727}$^{\hyperlink{aff6}{6},\hyperlink{aff5}{5}}$,
\orcidsymb{Khaled Said}{0000-0002-1809-6325}$^{\hyperlink{aff7}{7}}$,
\orcidsymb{Jesse van de Sande}{0000-0003-2552-0021}$^{\hyperlink{aff6}{6},\hyperlink{aff5}{5}}$,
\orcidsymb{Joss Bland-Hawthorn}{0000-0001-7516-4016}$^{\hyperlink{aff6}{6},\hyperlink{aff5}{5}}$,
\orcidsymb{Julia J. Bryant}{0000-0003-1627-9301}$^{\hyperlink{aff6}{6},\hyperlink{aff8}{8},\hyperlink{aff5}{5}}$,
\orcidsymb{Scott M. Croom}{0000-0003-2880-9197}$^{\hyperlink{aff6}{6},\hyperlink{aff5}{5}}$,
\orcidsymb{\'Angel~R.~L\'opez-S\'anchez}{0000-0001-8083-8046}$^{\hyperlink{aff9}{9},\hyperlink{aff10}{10},\hyperlink{aff5}{5}}$,
\orcidsymb{Nuria P. F. Lorente}{0000-0003-0450-4807}$^{\hyperlink{aff11}{11}}$,
\orcidsymb{Roberto Maiolino}{0000-0002-4985-3819}$^{\hyperlink{aff1}{1},\hyperlink{aff2}{2},\hyperlink{aff12}{12}}$ and
\orcidsymb{Edward N. Taylor}{0000-0002-5522-9107}$^{\hyperlink{aff13}{13}}$
}\vspace{0.4cm}
\\
\parbox{\textwidth}{
\hypertarget{aff1}{$^{1}$}Kavli Institute for Cosmology, University of Cambridge, Madingley Road, Cambridge, CB3 0HA, United Kingdom\\
\hypertarget{aff2}{$^{2}$}Cavendish Laboratory - Astrophysics Group, University of Cambridge, 19 JJ Thomson Avenue, Cambridge, CB3 0HE, United Kingdom\\
\hypertarget{aff3}{$^{3}$}Sterrenkundig Observatorium, Universiteit Gent, Krijgslaan 281 S9, B-9000 Gent, Belgium\\
\hypertarget{aff4}{$^{4}$}Research School of Astronomy and Astrophysics, Australian National University, Canberra, ACT 2611, Australia\\
\hypertarget{aff5}{$^{5}$}ARC Centre of Excellence for All Sky Astrophysics in 3 Dimensions (ASTRO 3D), Australia\\
\hypertarget{aff6}{$^{6}$}Sydney Institute for Astronomy, School of Physics, The University of Sydney, NSW, 2006, Australia\\
\hypertarget{aff7}{$^{7}$}School of Mathematics and Physics, The University of Queensland, Brisbane, QLD 4072, Australia\\
\hypertarget{aff8}{$^{8}$}Australian Astronomical Optics, Astralis-USydney, School of Physics, University of Sydney, NSW 2006, Australia\\
\hypertarget{aff9}{$^{9}$}School of Mathematical and Physical Sciences, Macquarie University, NSW 2109, Australia\\
\hypertarget{aff10}{$^{10}$}Macquarie University Research Centre for Astrophysics and Space Technologies,  NSW 2109, Australia\\
\hypertarget{aff11}{$^{11}$}AAO-MQ, Faculty of Science \& Engineering, Macquarie University. 105 Delhi Rd, North Ryde, NSW 2113, Australia\\
\hypertarget{aff12}{$^{12}$}Department of Physics and Astronomy, University College London, Gower Street, London WC1E 6BT, UK\\
\hypertarget{aff13}{$^{13}$}Centre for Astrophysics and Supercomputing, Swinburne University of Technology, Hawthorn, VIC 3122, Australia\\
}
}

\date{Accepted 2024 June 21. Received 2024 May 30; in original form 2023 July 19}

\pubyear{2024}

\begin{document}
\label{firstpage}
\pagerange{\pageref{firstpage}--\pageref{lastpage}}
\maketitle

\begin{abstract}
We use deep spectroscopy from the SAMI Galaxy Survey to explore the precision of the fundamental plane of early-type galaxies (FP) as a distance indicator for future single-fibre spectroscopy surveys. We study the optimal trade-off between sample size and signal-to-noise ratio (SNR), and investigate which additional observables can be used to construct hyperplanes with smaller intrinsic scatter than the FP.
We add increasing levels of random noise (parametrised as effective exposure time) to the SAMI spectra to study the effect of increasing measurement
uncertainties on the FP- and hyperplane-inferred distances.
We find that, using direct-fit methods, the values of the FP and hyperplane best-fit coefficients depend on the spectral SNR, and reach asymptotic values for a mean $\langle SNR \rangle =40\,\AA~\!\!\!^{-1}$.
As additional variables for the FP we consider three stellar-population observables: light-weighted age,
stellar mass-to-light ratio and a novel combination of Lick indices (\spindex).
For a $\langle SNR \rangle=45~\AA~\!\!\!^{-1}$ (equivalent to 1-hour exposure on a 4-m telescope), all three hyperplanes outperform the FP as distance
indicators. Being an empirical spectral index, \spindex avoids the
model-dependent uncertainties and bias underlying age and mass-to-light ratio
measurements, yet yields a 10~per\ cent reduction of the median distance uncertainty
compared to the FP.
We also find that, as a by-product, the \spindex hyperplane removes most
of the reported environment bias of the FP.
After accounting for the different signal-to-noise ratio, these conclusions also apply to a 50 times larger sample from SDSS-III. However, in this case, only $age$ removes the environment bias.
\end{abstract}



\section{Introduction}

The `fundamental plane' is an empirical multilinear relation between three galaxy
observables: size, velocity dispersion, and surface brightness
\citep[FP;][]{djorgovski+davis1987, dressler+1987}.
The relation deserved the name `plane' because of its remarkable tightness
(root mean square $rms\approx$\ 0.1~dex) relative to the range of the three
observables, which span $\approx$\ 0.5--1~dex. Nevertheless, the FP is known to
be warped \citep[e.g.,][]{yoon+park2022} and does not extend into the domain of
dwarf galaxies \citep{eftekhari+2022}. Moreover, the FP has non-zero
intrinsic scatter \citetext{\sigintr; \citealp{jorgensen+1996},
\citealp{hyde+bernardi2009}, \citealp{magoulas+2012}; hereafter:
\citetalias{magoulas+2012}}.

Because the FP can express distance-dependent size as a function of
distance-independent velocity dispersion and surface brightness, it can
be used as a distance indicator \citep[e.g.][]{hudson+1997, hudson+1999,
colless+2001}. Initially, the limiting factor of FP-derived distances were the
precision of the measured galaxy observables and small sample sizes. However,
the next generation of surveys was able to obtain more precise measurements
for larger samples \citetext{e.g., 6dFGS, the 6-degree Field Galaxy Survey,
\citealp{jones+2004}; and SDSS, the Sloan Digital Sky Survey, \citealp{york+2000}}.
For this new generation of surveys---and for every FP survey since---FP
cosmology was limited by the plane's \sigintr
\citetext{\citetalias{magoulas+2012}, \citealp{johnson+2014}, \citealp{said+2020}}.
This is a physical boundary that cannot be overcome by increasing the size of the sample, 
but requires understanding the physical drivers of \sigintr, be it new models (e.g.,
non-linear formulations of the `FP') or new observables (e.g., a higher-dimensional `FP').

Several studies have attempted to `explain' \sigintr; that is, to find the physical
origin of this scatter. The underlying hypothesis is that the FP is rooted in
the virial theorem, which links size and velocity dispersion to dynamical mass
\citep[e.g.][]{faber+1987, dressler+1987}. Going from the virial theorem to the 
FP means going from unobservable dynamical mass to luminosity, which we can
measure. This requires adding various terms that depend on the structure
and content of galaxies, but that can be expressed as functions of the FP
observables \citep[e.g.][]{prugniel+simien1996, prugniel+simien1997, graves+2009a}.
Unlike the virial theorem, these additional terms are expected to possess some
intrinsic scatter, due, e.g., to differences in the structure, mass distribution
and stellar-population content of galaxies, and it is in this physical scatter 
that the FP \sigintr likely originates. There is an extensive literature
concerning the physical origin of \sigintr, with various works pointing to
trends in the stellar-population properties \citep{gregg1992,
prugniel+simien1996, forbes+1998, graves+2009a, falcon-barroso+2011,
springob+2012, yoon+park2020}, including variations in the initial mass function
\citep[IMF;][]{graves+faber2010}, dark matter fraction \citep{zaritsky+2006}, or
structural variations \citep{prugniel+simien1997}.

The advent of extensive integral-field spectroscopy (IFS) surveys, with their large
collecting areas, has enabled observations of unprecedented depth for statistical
samples. Using a mass- and volume-limited sample from the SAMI Galaxy
Survey \citep{croom+2012}, \citet[][hereafter: \citetalias{deugenio+2021}]{deugenio+2021}
have shown that the residuals of the FP have the largest magnitude and most
significant correlation with stellar population age; in particular, at any
fixed position on the FP, the oldest/youngest galaxies lie on average below/above
the plane. Regardless of the physical interpretation of this result, this
suggests that age could be added to the FP as a fourth observable to remove the
age bias \textit{and} reduce \sigintr. While the problem of removing the age
bias has already been addressed \citep[e.g.,][]{guzman+lucey1993}, improving the
FP precision has proven more challenging.
Previous attempts were unsuccessful because the additional observational scatter
due to low-precision age measurements is greater than the prospective reduction in
\sigintr, producing an overall \textit{degradation} of the FP precision 
\citepalias{magoulas+2012}.
However, the quality of synthetic-aperture spectra from IFS surveys is much
better than that of even the best single-fibre surveys, enabling us to explore
the regime where age (or any other suitable observable) might be sufficiently
precisely determined to overcome both measurement uncertainties and intrinsic scatter
in the distance estimates.

Interestingly, and apparently unrelated, recent results using SDSS data suggest
that the FP suffers from environment bias, such that galaxies in rich/sparse environments
lie systematically below/above the plane \citetext{\citealp{howlett+2022}; hereafter:
\citetalias{howlett+2022}}. This could be a
problem for near-field cosmology applications of the FP, although the trend with
environment was found to be weak in \citetalias{deugenio+2021}. We note, however,
that age is known to correlate strongly with environment density, so that
the age and environment correlations with the FP residuals are likely related.

Any proposed reduction of \sigintr, to be advantageous to cosmology, must balance the 
requirement of higher data quality with sample size. For a survey of fixed total duration,
longer exposure times (required to lower \sigintr) come directly at the expense of smaller 
sample size (which increases the distance uncertainty). It is therefore reasonable to
expect that there is an optimal exposure time that minimises the \emph{total} distance 
uncertainty.

The goal of this work is twofold. First, we propose to realise the promise of a more
precise FP by folding in stellar population information of adequate precision;
this will in turn increase the accuracy of FP-derived distances, by
reducing the magnitude and significance of the residual correlation with environment.
Second, we also propose to find the optimal exposure time that a survey must adopt to
maximise the efficiency of the allocated total time.

The paper is structured as follows. In \S~\ref{s.das} we introduce the SAMI
Galaxy Survey, the measurements and the sample selection. We then compare the
FP to three alternative `hyperplanes', which add a stellar population age-related
observable as a fourth parameter (\S~\ref{s.rh}). We then discuss the implications of our
findings for future FP surveys (\S~\ref{s.d}) and conclude with a summary of
our findings (\S~\ref{s.c}).

Throughout this work, we assume a flat $\Lambda$CDM cosmology with $H_0 = 70 \;
\mathrm{km \; s^{-1} \; Mpc^{-1}}$ and $\Omega_m = 0.3$. All magnitudes are in
the AB system \citep{oke+gunn1983} and, unless otherwise specified, stellar
masses and mass-to-light ratios assume a Chabrier IMF \citep{chabrier2003}.


\section{Data and sample}\label{s.das}

The SAMI Galaxy Survey provides an unsurpassed combination of high
signal-to-noise ratio, large sample size and, crucially, a broad range of
environments, from isolated galaxies to dense galaxy clusters. The first goal
of this section is to present the data, referring the reader to the relevant
literature for a thorough description of the data reduction and verification
(\S~\ref{s.das.ss.sami}). The goal of this paper is to use the superior quality
of our IFS observations to study the effect of data quality on the precision of
FP-derived distances from single-fibre surveys. In \S~\ref{s.das.ss.degrade} we
describe the procedure we used to degrade the data quality, i.e. to simulate
observations with exposure times shorter than the actual exposure time of SAMI.
We then use these
simulated observations to measure aperture velocity dispersions
(\S~\ref{s.das.ss.kin.ext}), stellar population ages
(\S~\ref{s.das.ss.age.ext}) and a the new empirical index \spindex
(\S~\ref{s.das.ss.hfm.ext}).
In \S~\ref{s.das.ss.sample} we describe the sample selection criteria and,
finally, we illustrate the methods used to derive the best-fit parameters of the
4-d  hyperplane (\S~\ref{s.das.ss.methods}).

\subsection{The SAMI Galaxy Survey}\label{s.das.ss.sami}

\begin{figure*}
    \centering
    \includegraphics[width=\textwidth]{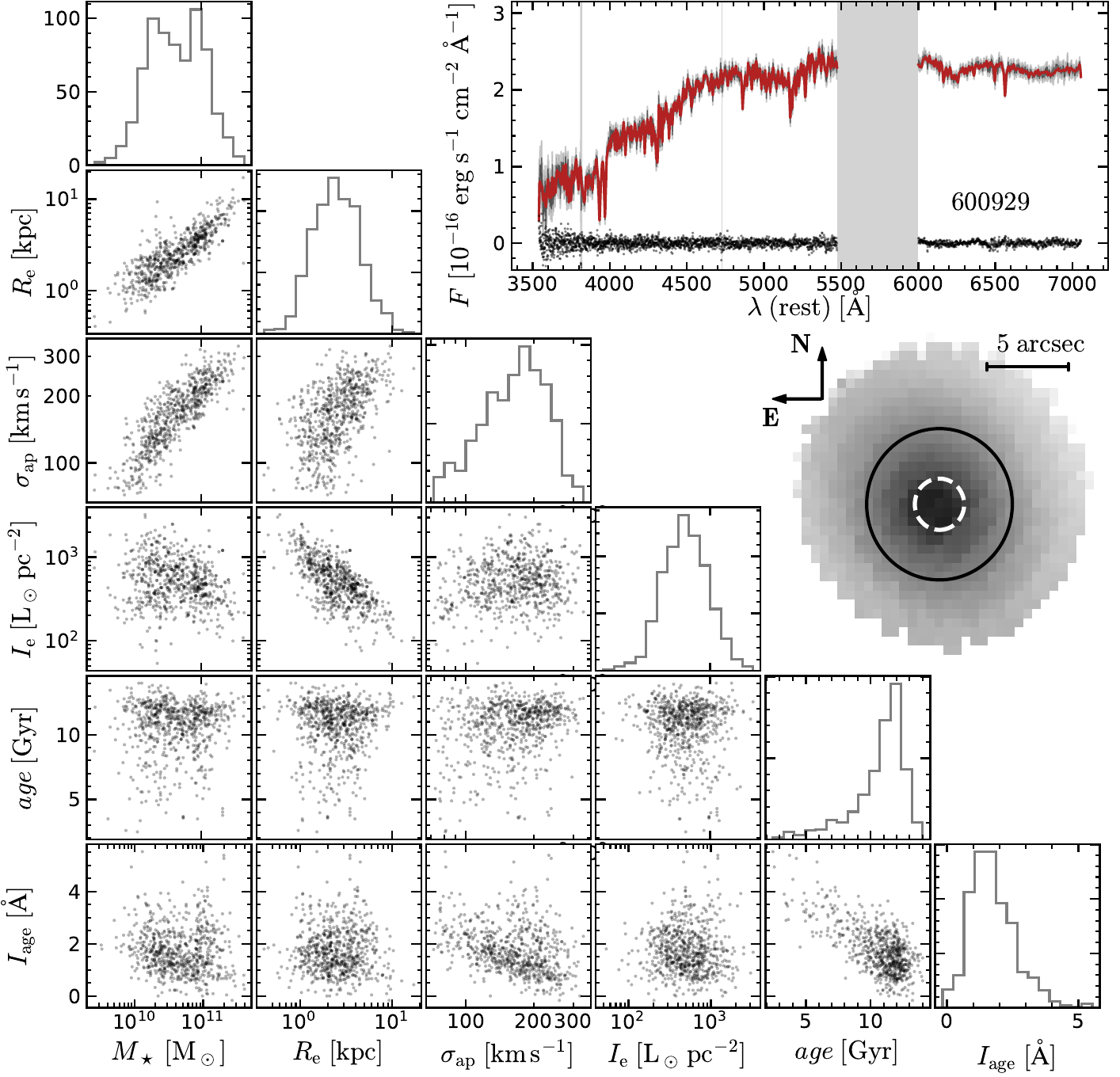}
    \caption{The properties of our sample of early-type galaxies, selected to
    mimic the SDSS sample of \citet{said+2020}. Note, in particular, the
    tail of galaxies with stellar population ages of 3--7~Gyr. These galaxies
    greatly increase the scatter of the FP (see Fig.~\ref{f.rh.plane.a}), but they
    cannot be rejected by selection on the \Halpha equivalent
    width. Besides, rejecting young early-type galaxies introduces a bias with
    environment, due to the link between local environmental density and
    stellar-population age (see Fig.~\ref{f.re.parcorr}). The \spindex spectral
    index, an empirical proxy for the FP residuals, is defined in
    \S~\ref{s.das.ss.hfm.ext}. The inset image is the
    synthetic IFU-derived image of galaxy 600929, which we randomly selected
    from the sample; the dashed white circle has a radius of 1.5~arcsec and denotes
    the synthetic aperture we used for our spectroscopy, while the solid
    black circle shows the half-light radius,
    $R_\mathrm{e}$ (4.1~arcsec). The top-right panel shows the spectrum within the 
    1.5-arcsec radius for the same galaxy; the black line is
    the data, the grey region above/below the data is the 1-\textsigma
    uncertainty, the red line is the {\sc ppxf} best-fit spectrum and the dots
    are the residuals (the vertical grey bands are not fit).
    }\label{f.ds.sample}
\end{figure*}

The SAMI Galaxy Survey (hereafter simply SAMI) is the Australian
state-of-the-art integral-field-spectroscopy extragalactic survey
\citep{croom+2012}. SAMI used the Sydney-AAO Multi-object Integral field
spectroscopy instrument (hereafter: the SAMI instrument), formerly mounted at the
prime focus of the 4-metre Anglo-Australian Telescope. The SAMI instrument is 
endowed with 13 independent
integral field units (IFUs), deployable anywhere inside a circular field of view
of diameter 1~degree. Each IFU consists of 61 fused fibres of 1.6-arcsec
diameter \citep[hexabundles;][]{bland-hawthorn+2011,
bryant+2014}: compared to ordinary fibre bundles, hexabundles are more fragile,
but have 50~per\ cent-larger fill factor \citep{croom+2012}.

The 13 IFUs, alongside 26 individual sky fibres, are plugged into a pre-drilled
metal plate, and are held in position using magnetic connectors. The fibres feed
the double-beam AAOmega spectrograph \citep{sharp+2006}, configured with the
570V grating at 3750--5750~\AA\ (blue arm) and the R1000 grating at 6300--7400~\AA\ 
(red arm). The spectral resolution of the blue arm is $R=1812$ ($\sigma = 70.3
\, \mathrm{km\, s^{-1}}$), whereas for the red arm it is
$R=4263$~\citep[$\sigma = 29.9 \, \mathrm{km\,s^{-1}}$;][]{vandesande+2017a}.
Each plate is observed for approximately 3.5~hours, using
seven dithered observations arranged in a hexagonal pattern
\citep{sharp+2015}. The median full-width-at-half-maximum (FWHM) seeing is
2.06~arcsec, with a standard deviation of 0.40~arcsec. We
use data available as part of the third public data release \citep[DR3;][]{
croom+2021a}. The data reduction is originally described in \citet{sharp+2015} and
\citet{allen+2015}, with a number of improvements explained in the subsequent
public data release papers \citep{green+2018, scott+2018, croom+2021a}.

To match spectroscopy from the SDSS-I and II surveys, we use 1-d spectra from the
synthetic circular apertures of radius 1.5~arcsec\footnote{
Note this is different from the SDSS-III spectra, which used the BOSS spectrograph
and fibres of 1-arcsec radius. However, the BOSS sample is not appropriate for near-field
FP cosmology and is not used here.
}, which are available in the
public \href{https://docs.datacentral.org.au/sami/}{SAMI database}. The spectra
from the blue and red arms of the spectrograph are combined into a single spectrum
of uniform spectral resolution \citep{vandesande+2017a}.

\subsection{Degrading the SAMI spectra}\label{s.das.ss.degrade}

The long exposure times of SAMI ensure a higher signal-to-noise ratio (SNR) than
that of typical single-fibre surveys (for our sample, we have a median rest-frame
$g-$band SNR of 86.7~\AA$^{-1}$, whereas the sample from \citetalias{howlett+2022} has
24.2~\AA$^{-1}$).
However, this superior data quality comes at the price of sample size, which for SAMI is
smaller by an order of magnitude. While we can reasonably predict that
increasing the sample size by a factor $N$ would decrease the uncertainty of the FP
parameters by $\sqrt{N}$, understanding the effect of data quality on the
uncertainty of age measurements is much more challenging, due to the non-linear
nature of the measurement. For this reason, we propose to simulate the effect of
\textit{shorter} exposure times on the quality of the SAMI spectra. To do so, we need
first a robust estimate of the actual SNR, which we obtain from the residuals of the best-fit
spectrum. We fit the SAMI aperture spectra with the MILES stellar template library
\citep{sanchez-blazquez+2006}, using
{\sc \href{https://pypi.org/project/ppxf/}{{\sc ppxf}}}
\citep[penalised PiXel Fitting,][]{cappellari+emsellem2004, cappellari2017,
cappellari2022} and
assuming a Gaussian line-of-sight velocity distribution (LOSVD).
Following \citet{vandesande+2017a}, we use a 12\textsuperscript{th}-order additive
Legendre polynomial.
We mask both strong sky lines and regions that are subject
to strong emission lines from the gas.
For SAMI, the instrument spectral resolution is uniform
\citep[full-width half-maximum (FWHM) of 2.65~\AA,][]{vandesande+2017a}; for the
MILES library, we use a uniform FWHM of 2.51~\AA\ \citep{falcon-barroso+2011}.
For most galaxies, SAMI spectral resolution is coarser than the MILES value, so
we match the two by broadening the templates with the appropriate kernel.
However, for 240 galaxies with redshift $z>0.0558$, the rest-frame
spectral resolution of SAMI is better than 2.51~\AA, so, in principle, we should
degrade the data to match the templates. To avoid this, we take no action, which
introduces a bias in the FWHM of up to 4~per\ cent for the magnitude-limited sample, at
the maximum redshift $z=0.0997$. However, the effect on the measured kinematics is
smaller than 4~per\ cent, because our sample has even higher physical dispersion (the
1\textsuperscript{st} percentile is 77 $\mathrm{km\,s^{-1}}$), and because, to
first order, the instrument spectral resolution is subtracted in quadrature from
the observed broadening.

From the residuals of the best-fit spectrum, we estimate the noise spectrum as
follows.
First, we calculate the biweight scale of the residuals in a moving window of
54~\AA\ (this is a robust estimate of the standard deviation). Masked pixels are
not considered
in the estimate. Afterwards, the noise vector $n_\mathrm{SAMI}(\lambda)$ is linearly
interpolated over missing pixels. The resulting `true' signal-to-noise ratio, $\mathrm{SNR_{SAMI}}(\lambda)$, is
defined as the ratio between the spectrum and $n_\mathrm{SAMI}(\lambda)$.
To simulate the spectrum from an exposure of duration \texp, we need to define a
noise spectrum from which to draw random noise, then add this noise to the original
SAMI spectrum. We first calculate the ratio
$f_\mathrm{exp}$ between \texp and the true exposure time \texp[SAMI]; we
then estimate the resulting $\mathrm{SNR}(\lambda)$ as $\sqrt{f_\mathrm{exp}} \,
\mathrm{SNR_{SAMI}}(\lambda)$.
The additional noise required to go from $\mathrm{SNR_{SAMI}}(\lambda)$ to the target
SNR is equal to
\begin{equation}
    n_\mathrm{SAMI}(\lambda) \,
    \sqrt{\left(\frac{\mathrm{SNR_{SAMI}}(\lambda)}{\mathrm{SNR}(\lambda)}\right)^2-1} ~.
\end{equation}
For each pixel, we draw randomly from the Gaussian distribution with the corresponding
noise, and add the result to the SAMI spectrum. The SAMI noise vector is updated by adding
in quadrature the additional noise.

For each galaxy in SAMI, we repeat this procedure for the following exposure times:
0.125, 0.1875, 0.25, 0.375, 0.5, 0.75, 1, 1.5, 2, and 3~hours, thus creating a set of ten
mock surveys. These exposure times are related by factors of 2 from 2~hours and 3~hours. Note
that whenever $\texp = \texp[SAMI]$, no noise is added. Moreover, whenever
$\texp>\texp[SAMI]$, that galaxy is excluded from the mock survey; in
practice, this happens only for the mocks with $\texp=3$~hours, because only 266
galaxies have $\texp[SAMI]<3$~hours (8~per\ cent), and none have $\texp[SAMI]<2$~hours.

\begin{figure}
    \centering
    \includegraphics[trim={0 0 0 0.7cm},clip,width=\columnwidth]{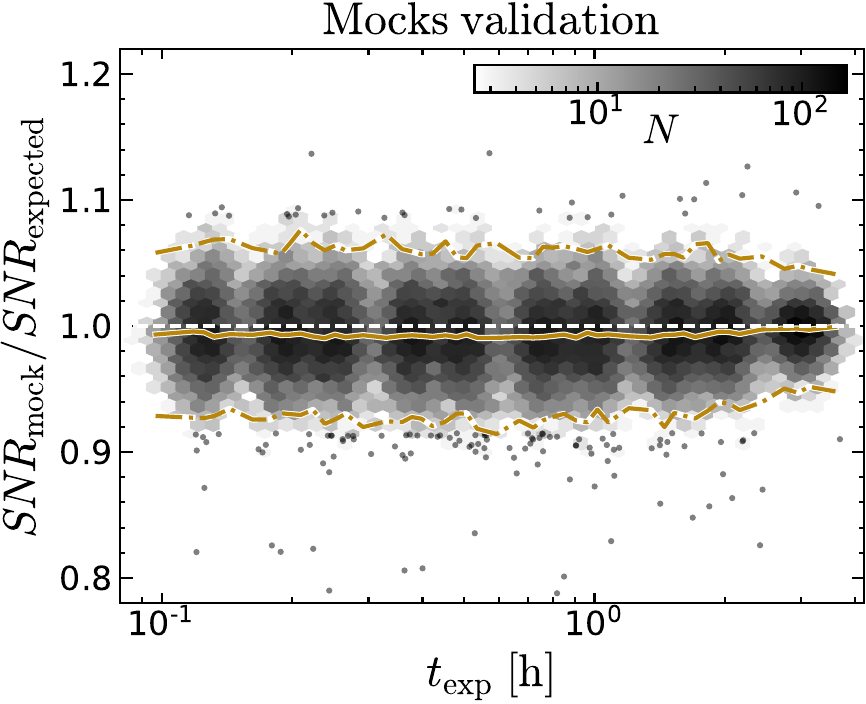}
    \caption{Ratio between the measured and expected SNR for all mock spectra,
    as a function of the mock exposure time; more than 99~per\ cent of the ratios are
    within 10~per\ cent. The expected SNR is given by the true SNR of the SAMI spectra,
    multiplied by the square root of the ratio between the mock exposure time
    and the SAMI exposure time; the measured SNR is the standard deviation of
    the residuals from the best-fit {\sc ppxf} spectrum to the mock spectrum.
    The white dashed line traces unity (the value expected from a
    self-consistent noise model); the solid and dot-dashed lines trace
    respectively the median, and the 
    1\textsuperscript{st}--99\textsuperscript{th} percentiles of the mocks.
    Bins represent regions with more than two mocks; individual mocks are
    represented by black dots.
    Note that there are only ten exposure times, but we randomly perturbed
    these discretised values by 10~per\ cent for display purposes.
    }\label{f.das.noise}
\end{figure}

After creating these mock observations, we measure their empirical SNR and, as
a sanity check, we verify that it falls close to the expected SNR. In Fig.~\ref{f.das.noise}, we compare
the empirical SNR measured from the degraded mock spectrum to the SNR
expected given the original SNR and exposure time.
The ratios are shown as a function of \texp, for each of the ten mocks
(for display purposes, the exposure times of individual mock spectra have
been randomly perturbed by 10~per\ cent).
The ratio of SNRs are normalised by the square root of the ratio of the
exposure times, so the values should be close to unity (white dashed line in
Fig.~\ref{f.das.noise}). Indeed, only 0.2~per\ cent
of the mocks have a SNRs ratio more than 10~per\ cent from unity, only 0.03~per\ cent
have discrepancy more than 20~per\ cent from unity, and none lie beyond 30~per\ cent from
unity. There is a bias in that the mocks' SNR is lower than the expected
value, with the median ratio equal to 0.993 (solid line in
Fig.~\ref{f.das.noise}). In practice, this bias is negligible compared to real
effects that we do not model, e.g., the fact that at some point, noise due to sky
subtraction residuals or even read-out noise may become dominant over photon noise,
invalidating our assumed scaling of the SNR with the square root of the exposure time.
We conclude that the SNR estimates and the degrading pipeline work as
expected.

\subsection{Sample \texorpdfstring{$SNR$}{\textit{SNR}} vs exposure time}\label{s.das.ss.snr}

\begin{figure}
    \centering
    \includegraphics[trim={0 0 0 0.9cm},clip,width=\columnwidth]{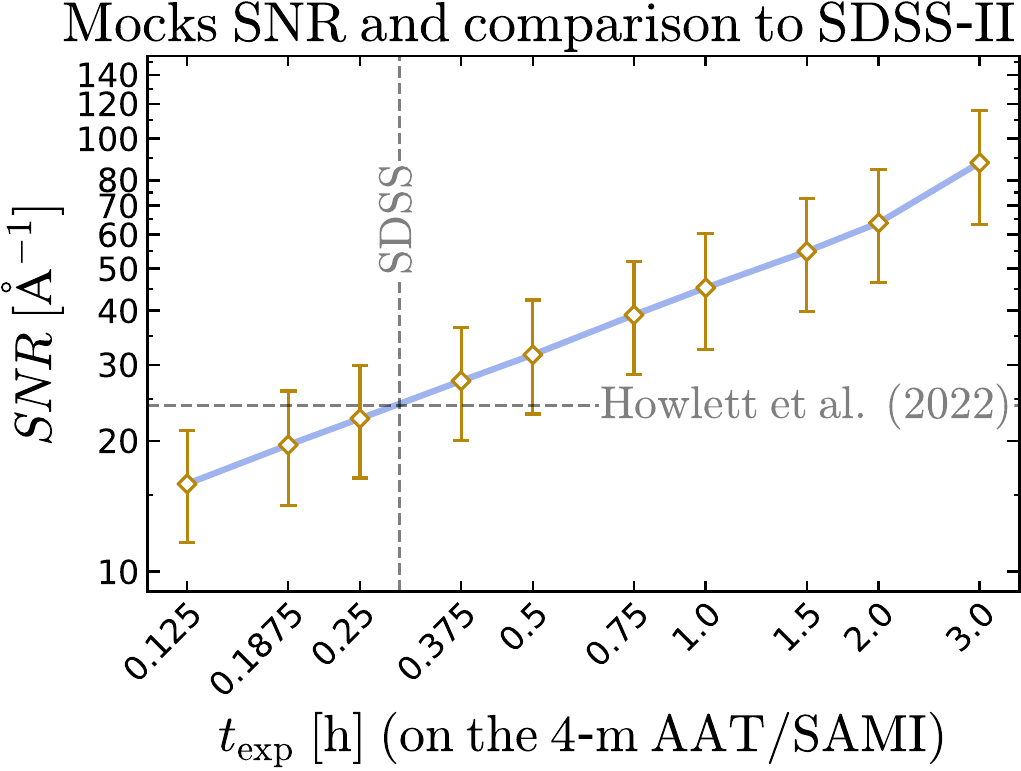}
    \caption{Median sample $g-$band $SNR$ versus mock exposure time. The
    errorbars mark the 16\textsuperscript{th}--84\textsuperscript{th}
    percentiles of the $SNR$ distribution for each mock; the dashed lines
    are the exposure time of SDSS (rescaled to AAT/SAMI) and the
    median $SNR$ of the SDSS sample from \citetalias{howlett+2022}; interpolating
    between the mocks we predict SDSS to have a median $SNR$ very close to
    the experimental value.
    }\label{f.das.snr}
\end{figure}

A key validation of our mock data is the relation between \texp and $SNR$.
To each value of \texp, we associate $\langle SNR \rangle$, the rest-frame
$g-$band empirical $SNR$ in units of rest-frame \AA$^{-1}$, calculated as the
median over the whole sample of 703 galaxies. The results are shown
in Fig.~\ref{f.das.snr}, where for each of the ten mocks the diamonds represent
$\langle SNR \rangle$ and the errorbars encompass the
16\textsuperscript{th}--84\textsuperscript{th} percentiles of the $SNR$
distribution. As a benchmark, we report the equivalent exposure time of SDSS 
spectroscopy (\texp[SDSS], vertical dashed line). We assume the total
throughput of SDSS and AAT/SAMI to be comparable, so we estimate
\texp[SDSS] on AAT/SAMI by rescaling the true SDSS exposure time of
0.75~hours by the ratio of the telescope areas and get
$\texp[SDSS] = 0.3$~hours. We already reported that the $g-$band
$\langle SNR \rangle$ of the SDSS sample from \citetalias{howlett+2022} is
24.2~\AA$^{-1}$ (horizontal dashed line in Fig.~\ref{f.das.snr}; see also
\S~\ref{s.rs}). As an independent validation of our mocks, the solid
blue line (which simply connects the diamonds to guide the eye) passes very
close to where the dashed lines $\texp=0.3$~hours and
$\langle SNR \rangle = 24.2$~\AA$^{-1}$ intersect. Interpolating linearly
between the mocks with $\texp = 0.25$ and $0.375$~hours, which have
$\langle SNR \rangle$ of 22.5 and 27.5~\AA$^{-1}$, we find that for
\texp[SDSS] we would obtain $\langle SNR \rangle = 25.5$~\AA$^{-1}$, only 
5 per~cent larger than the experimental value.

\subsection{Aperture velocity dispersions}\label{s.das.ss.kin.ext}

\begin{figure}
    \centering
    \includegraphics[trim={0 0 0 2.2cm},clip,width=\columnwidth]{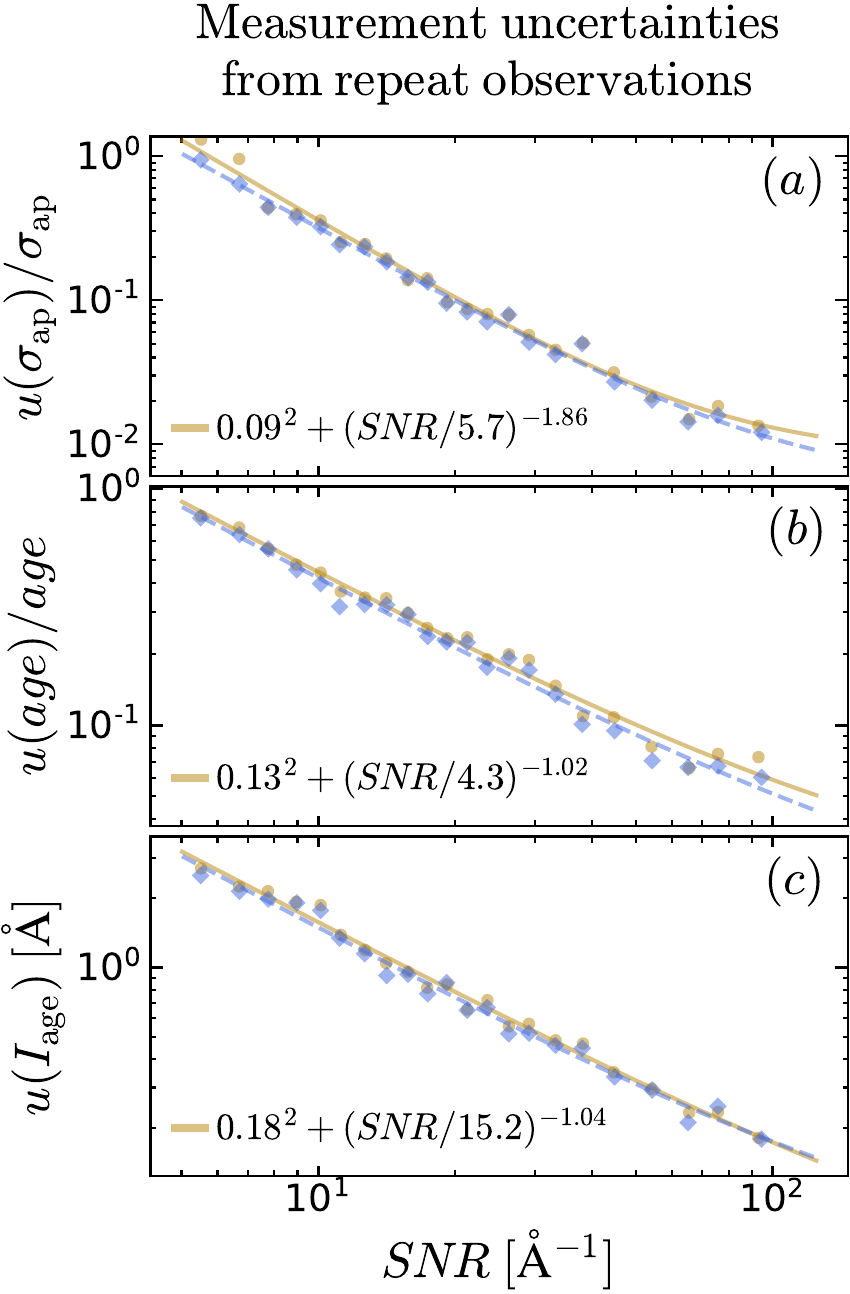}
    {\phantomsubcaption\label{f.das.uncert.a}
     \phantomsubcaption\label{f.das.uncert.b}
     \phantomsubcaption\label{f.das.uncert.c}}
    \caption{Measurement uncertainties from repeat observations, as a function
    of empirical SNR, for \sigap (panel~\subref{f.das.uncert.a}),
    $age$ (panel~\subref{f.das.uncert.b}) and the empirical index \spindex
    (panel~\subref{f.das.uncert.c}). The golden circles are measurement
    uncertainties derived from repeat observations; these are obtained by
    binning the difference between repeats in SNR, removing pairs where the
    two $SNR$s differ by more than 50~per\ cent, then taking the standard
    deviation of the difference in each bin (divided by $\sqrt{2}$). The solid
    golden lines are best-fit relations, which we use to estimate the
    measurement uncertainty on our data. Blue diamonds/dashed lines are the
    same, but rejecting pairs where the $SNR$s differ by more than 10~per\ cent.
    These uncertainties do not account for systematic errors (see e.g.\ 
    Fig.~\ref{f.das.agecom}), but systematics are not relevant to our analysis.
    }
    \label{f.das.uncert}
\end{figure}

Aperture velocity dispersions \sigap are measured with the same algorithm we
applied to get the best fit used in the SNR estimate. The data are validated by comparing
these measurements to the corresponding measurements from \citetalias{deugenio+2021}.
We find 
\begin{equation}
    \log \sigap = (0.962\pm0.004)\log\sigma_\mathrm{ap,DE21} + (0.086\pm0.009)
\end{equation}
with a scatter of 0.016~dex, fully explained by measurement uncertainties (see
below).
To estimate the measurement uncertainties, we use repeat observations. We have
overall 165 galaxies with two observations, 62 with three observations,
and 21 galaxies with four observations. These can be arranged in 477 possible
pairs,
of which we consider only 352 independent pairs. Considering together all ten
mock samples, we have 3520 pairs of spectra.
We reject 330 pairs where the SNRs differ by more than 50~per\ cent (using a stricter cut
of 10~per\ cent removes 2073 pairs, but does not change our results, as described below).
To estimate the relative uncertainties, we calculate the natural logarithm of the
ratio between each pair of \sigap measurements, then take the biweight scale of
the logarithm of the ratio in twenty-one bins in SNR (changing the number of
bins does not change the final estimated uncertainties). The result is shown by
the golden circles in Fig.~\ref{f.das.uncert.a}. We fit these twenty-one values
and find that the uncertainty about \sigap can be expressed as
\begin{equation}\label{eq.das.uncert.sig}
    \frac{u(\sigap)}{\sigap} = (0.09\pm0.01)^2 + \left( \frac{\mathrm{SNR} \; \mathrm{[\AA^{-1}]}}{5.7\pm0.3}\right)^{-(1.86\pm 0.12)}
\end{equation}
(solid golden line in Fig.~\ref{f.das.uncert.a}).
The uncertainties about the best-fit parameters of Eq.~(\ref{eq.das.uncert.sig}) are
estimated by bootstrapping the sample of 352 pairs of measurements one thousand times,
then taking the biweight scale of the distribution of each of the three parameters.
If we were to reject all pairs where the two SNRs differ by more than 10~per\ cent, we
would find the same result, as highlighted by the blue diamonds in
Fig.~\ref{f.das.uncert.a}; in particular, the best-fit parameters of
Eq.~(\ref{eq.das.uncert.sig}) would be $0.08\pm0.02$, $5.1\pm0.4$ and
$-1.76\pm0.11$ (dashed blue line in Fig.~\ref{f.das.uncert.a}). From here on,
the uncertainties on \sigap are estimated using Eq.~(\ref{eq.das.uncert.sig}).

\subsection{Light-weighted stellar population ages}\label{s.das.ss.age.ext}

We use light-weighted stellar-population age from full spectral fitting (FSF),
obtained again from {\sc ppxf}. As input spectra, we use the MILES simple
stellar population (SSP) library \citep{vazdekis+2010, vazdekis+2015}, with
BaSTI isochrones \citep{pietrinferni+2004, pietrinferni+2006} and solar
abundance. The age-metallicity grid consists of 53 age bins
spanning 0.03~Gyr~$< \mathrm{SSP\;age} <$~14~Gyr and 12 metallicity bins spanning
$-2.27 < [Z/H] < 0.4$.
To match the spectral resolution of this library to the SAMI data, we use the same
procedure adopted in \S~\ref{s.das.ss.degrade}, because the MILES SSP has the same
spectral resolution as the MILES stellar template library used in
\S~\ref{s.das.ss.degrade}. This time, we use multiplicative Legendre polynomials,
following \citet{owers+2019}. No regularisation is applied because we prioritise
goodness of fit over realistic star-formation histories. The light-weighted stellar
population age is averaged in log-space \citep{mcdermid+2015}. From here on, we
refer to these ages measurements in italics ($age$); when we want to refer
to age in general, we use normal font `age'.

The uncertainties on $age$ are estimated
as for \sigap, by comparing repeat observations. Fig.~\ref{f.das.uncert.b} shows
the value of the uncertainty on $age$ as a function of SNR; the
symbols have the same meaning as for \sigap (\S~\ref{s.das.ss.age.ext}). For individual
spectra of a given SNR, we calculate the relative uncertainty on $age$
\begin{equation}\label{eq.das.uncert.age}
    \frac{u(age)}{age} = (0.13\pm0.08)^2 + \left( \frac{\mathrm{SNR} \; \mathrm{[\AA^{-1}]}}{4.3\pm0.3}\right)^{-(1.02\pm0.07)}
\end{equation}
(solid golden line in Fig.~\ref{f.das.uncert.b}).
Using the stricter cut at 10~per\ cent, we would get somewhat smaller uncertainties
(up to 20~per\ cent smaller at the high-SNR end, cf.\ the solid golden and dashed blue
lines in Fig.~\ref{f.das.uncert.b}). In this case, the best-fit relation
describing the uncertainties has parameters: $0.12\pm0.07$, $4.3\pm0.2$ and
$-1.05\pm0.07$; as can be seen, all parameters, taken singly, are
statistically consistent with Eq.~(\ref{eq.das.uncert.age}). From here on,
we use the more conservative estimate of the
uncertainties on $age$ as inferred from Eq.~(\ref{eq.das.uncert.age}).

\begin{figure}
    \centering
    \includegraphics[width=\columnwidth]{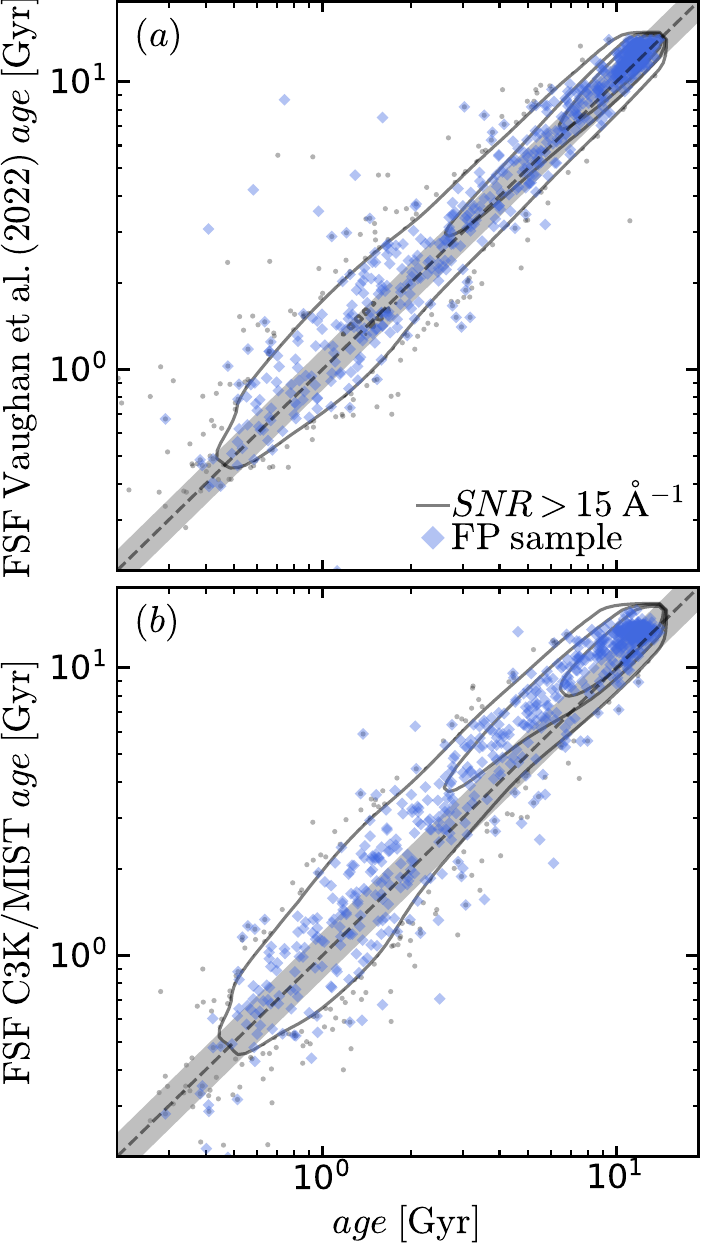}
    {\phantomsubcaption\label{f.das.agecom.a}
     \phantomsubcaption\label{f.das.agecom.b}}
    \caption{Comparison of our default age measurements (from full spectral fitting,
    using the MILES SSP library, labelled $age$) to two
    alternative age measurements: the measurements obtained with the methodology
    fo \citet[][panel~\subref{f.das.agecom.a}]{vaughan+2022}, and the measurements
    obtained using a different SSP library (panel~\subref{f.das.agecom.b}). The
    contours enclose the 30\textsuperscript{th}, 50\textsuperscript{th} and
    90\textsuperscript{th}
    percentiles of the distribution of all galaxies with $SNR > 15 \,
    \AA^{-1}$
    (galaxies outside the 90\textsuperscript{th}-percentile contour are
    represented by individual grey dots); the blue diamonds are the 703 FP
    galaxies from the mock with
    $\texp = 2$~hours. The strong systematics between the different
    age measurements underscore the importance of assumptions in measuring age,
    but we checked that our results do not depend on these assumptions.
    }\label{f.das.agecom}
\end{figure}

In Fig.~\ref{f.das.agecom.a} we compare our default ages to the measurements
from \citet{vaughan+2022}, also obtained using {\sc ppxf} and the MILES SSP libarry.
Grey dots are galaxies with $SNR>$15~\AA$^{-1}$ and blue diamonds are galaxies
from the FP sample (see \S~\ref{s.das.ss.sample}). Note that 15~\AA$^{-1}$ is
the 5\textsuperscript{th} percentile of the SNR distribution of the FP sample,
but the grey circles include both star-forming and quiescent galaxies.
We further validate our results by comparing our
default $age$ measurements to the values obtained by replacing the MILES SSP
library with the synthetic C3K library \citep{conroy+2019} with
MIST isochrones \citep{choi+2016}. The advantage of
this synthetic library is that it has uniform coverage in age-metallicity space,
whereas MILES has limited coverage over the grid we use\footnote{Please refer to
the \href{http://research.iac.es/proyecto/miles/pages/ssp-models/safe-ranges.php}{MILES webpage}}.
The C3K/MIST library has an age-metallicity grid consisting of 107 age bins with
0.1~Myr~$< \mathrm{SSP\;age} <$~20~Gyr (uniformly spaced in log-age by
0.05~dex) and 12 metallicity bins with $-2.5<[Z/H]<0.5$ (uniformly spaced
by 0.5~dex). For our test, we removed all SSP spectra with SSP age $<0.03$~Gyr,
to match the age range of MILES. Despite this, differences between the
grid of the two SSP libraries still persist, both in the metallicity grid and in
the spacing of the 57 age bins. In Fig.~\ref{f.das.agecom.b} we compare our
default age measurements to the values obtained using the (trimmed) C3K/MIST library
(labelled `FSF~C3K/MIST $age$').

We have checked that replacing $age$ with the measurements from \citet{vaughan+2022}
or FSF~C3K/MIST~$age$ in our analysis does not alter our general conclusions.
We also tested our results by using the Lick-index based SSP-equivalent ages 
\citep{scott+2017}; even though these measurements are very different from their FSF
equivalents \citep[see e.g,][]{mcdermid+2015}, our conclusions are not affected.

\begin{figure*}
    \includegraphics[width=\textwidth]{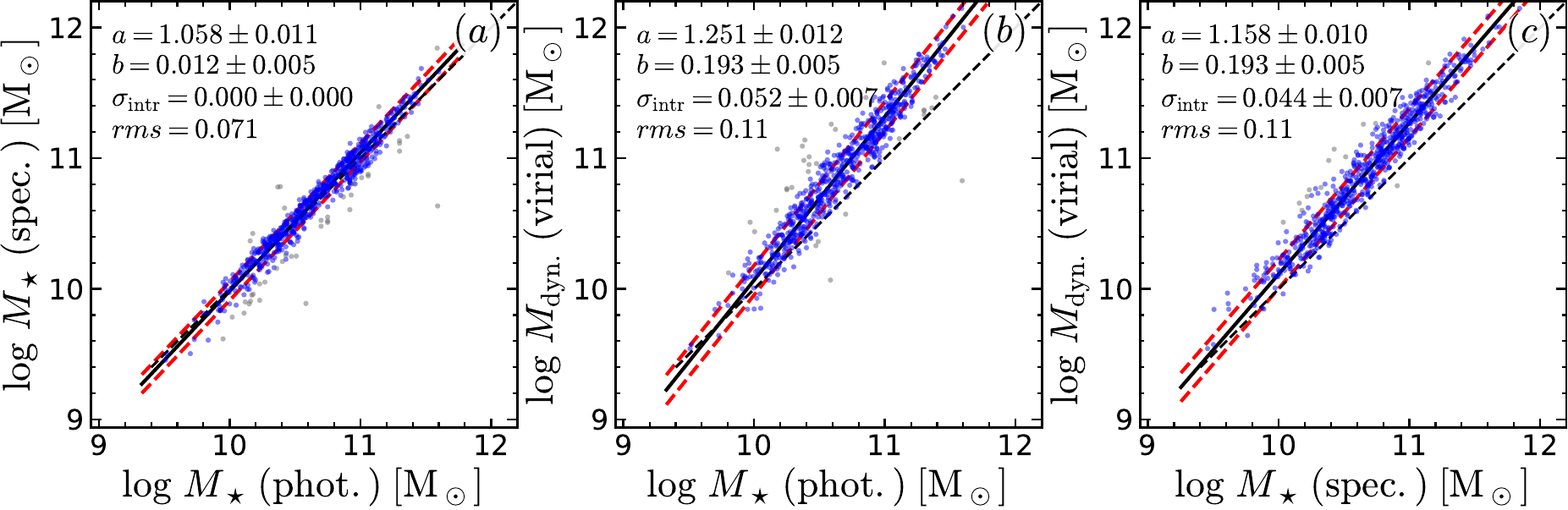}
    {\phantomsubcaption\label{f.das.ml.a}
     \phantomsubcaption\label{f.das.ml.b}
     \phantomsubcaption\label{f.das.ml.c}}
    \caption{Comparison between photometric and spectroscopic $M_\star$ (panel~\subref{f.das.ml.a}, and their scaling with virial mass (panels~\subref{f.das.ml.b} and~\subref{f.das.ml.c}), which we use as a benchmark. Surprisingly, panels~\subref{f.das.ml.b} and~\subref{f.das.ml.c} have the same $rms$, which implies that
    $M_\star$(phot.) is as precise as $M_\star$(spec.). This could be due to $M_\star$(phot.) using information at longer wavelengths ($i$-band) which is not covered by SAMI spectroscopy.}\label{f.das.ml}
\end{figure*}

In addition to $age$, we also retrieve the $r$-band stellar mass-to-light ratio,
$\Upsilon_\star$. This gives us two stellar-mass measurements, the default $M_\star$ calibrated from multi-band photometry and $i$-band absolute magnitude \citep{taylor+2011},
and a new $M_\star$ from $g$- and $r$-band spectroscopy and $r$-band absolute magnitude.
The two measurements are compared in Fig.~\ref{f.das.ml.a}, showing good agreement. The Spearman correlation coefficient between the two measurements of mass-to-light ratio is $\rho=0.5$, with the probability $P$ of no correlation being $P<10^{-10}$.

We use virial mass from \citet{cappellari+2013a} as a benchmark, and find that photometric stellar masses give the same observed scatter as spectroscopic stellar masses (panels~\subref{f.das.ml.b} and~\subref{f.das.ml.c}). This is surprising, because spectroscopy gives access to much more information than photometry alone.
This unexpected outcome could be due to photometric $M_\star$
using $i$-band photometry, which is redder than the $r$-band spectroscopy available in SAMI.
A similar result was found by \citet{hyde+bernardi2009} using SDSS data.
Overall, we find that the stellar-mass plane \citep{hyde+bernardi2009} does not give a more precise scaling relation, even when based on photometric $M_\star$ (Appendix~\ref{app.mstarplane}).

\subsection{A new empirical index}\label{s.das.ss.hfm.ext}

Our declared aim is to realise the gain in distance precision expected from
adding stellar-population age to the FP. However, even if this attempt was
successful, it retains a critical flaw because of how we measure the age of
a galaxy's stellar population. A comparison with \sigap will illustrate the
problem. To measure \sigap, we use empirical stellar spectra and assume a
Gaussian LOSVD. This might present problems if the library of stellar spectra
(derived from the Milky Way) cannot reproduce the spectra of external galaxies,
or if the LOSVD is not Gaussian. In practice, however, Milky Way based stellar
templates reproduce very accurately most of the visible spectrum for all but the
most massive galaxies \citep[e.g.][their fig.~25]{vandesande+2017a}. Moreover,
within at least 1~$R_\mathrm{e}$, LOSVD deviations from a Gaussian are small
(e.g. D'Eugenio et~al., in~prep.). Past experience shows that comparing between different
methods gives consistency to within a modest fixed scale error \citep[e.g.][]{said+2020}.
In contrast, stellar-population age measurements are either based on empirical
stellar-population spectra (derived from Milky Way stars), or on theoretical
spectra (which are difficult to calibrate in the age and chemical-abundance
ranges where there are few or no Milky Way stars). This is often compounded by
assumptions about the initial mass function and the star-formation history, as
well as by the degeneracy between flux calibration uncertainties and the effects
of age and dust. Full spectral fitting methods attempt to reproduce the whole
optical spectrum, while methods based on spectral indices `condense' the spectral
information around specific spectral regions. Finally, some authors combine
optical spectroscopy with broad-band photometry at bluer and redder wavelengths.
In this broad and entangled landscape, age measurements are dominated by (non-linear)
systematics, a polite way to state that often the measurements are driven by the
method as much as the data. Fig.~\ref{f.das.agecom} shows an example of the
magnitude of systematic age differences.

\begin{figure}
    \centering
    \includegraphics[width=\columnwidth]{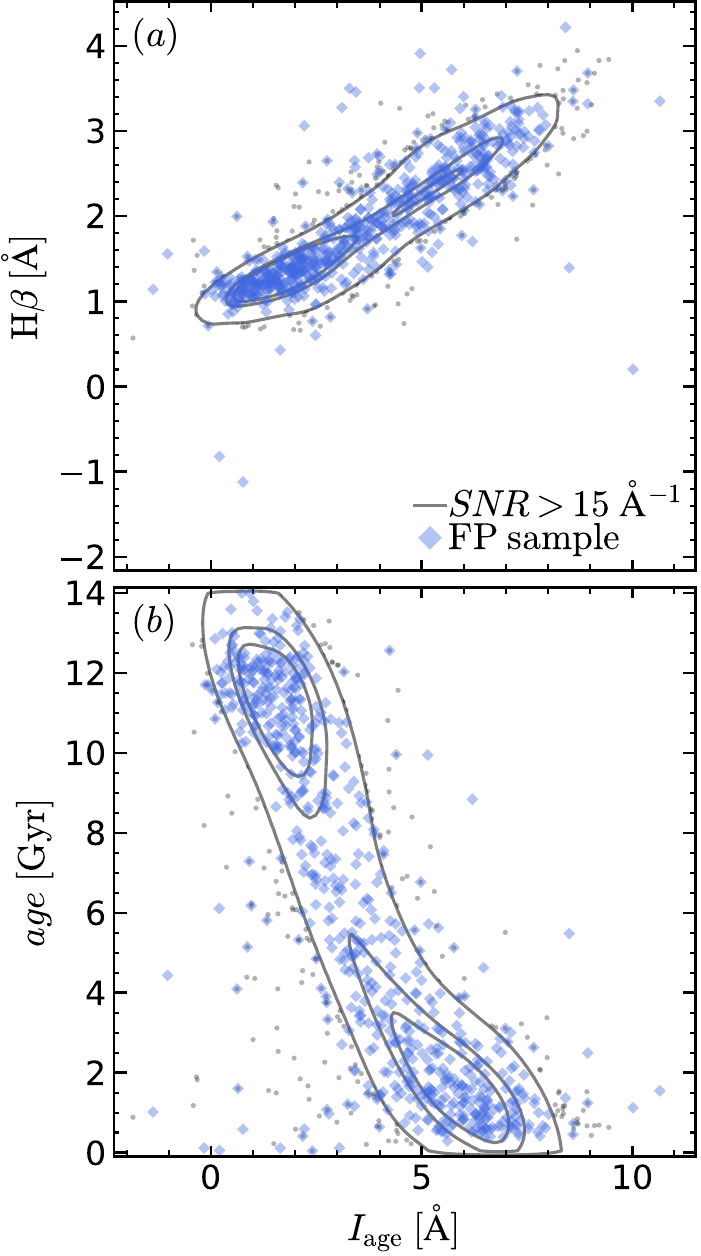}
    {\phantomsubcaption\label{f.das.emlicom.a}
     \phantomsubcaption\label{f.das.emlicom.b}}
    \caption{
    Comparing the index \spindex to the $\mathrm{H}\beta$ Lick index
    (panel~\subref{f.das.emlicom.a}) and to $age$ (panel~\subref{f.das.emlicom.b};
    all symbols are the same as in Fig.~\ref{f.das.agecom}). \spindex is
    tightly correlated to $\mathrm{H}\beta$ and $age$; the correlation with
    $age$ means that, just like $age$, \spindex can be used to predict the FP
   residuals (see \S~\ref{s.rh.ss.planes}, cf.\ Fig.~\ref{f.rh.plane.b}
   and \subref{f.rh.plane.c}).
    }\label{f.das.emlicom}
\end{figure}

It is highly undesirable for a distance estimator to be dominated by
systematics, because it hinders comparison between different works. For this
reason, we define a new empirical index \spindex to replace age in the FP.
This index is the sum of five Lick indices
\begin{equation}\label{eq.das.spindex}
    I_{\mathrm age} \equiv {\rm H\text{\textdelta}_F + H\text{\textbeta} + Fe5015 - Mgb - Fe5406}
\end{equation}
This index was derived empirically, by studying the sign and magnitude of the correlations
between the FP residuals and the set of classic Lick indices. We did not attempt a more
systematic search, such as `sequential feature selection' \citetext{\citealp{ferri+1994}; e.g., \citealp{davis+2022}}.
Our current goal is to demonstrate a purely empirical measurement that could aid reducing the scatter of the FP, which we will demonstrate in Section~\ref{s.rh}.
A more systematic search is left for future work.
Note that we measure the indices at the native resolution of the SDSS data
(including kinematic line broadening), without matching the spectral resolution
of the original Lick definition. Thus our (pseudo) Lick indices contain both
stellar population and stellar kinematics information.
The measurement uncertainties on \spindex are derived from repeat observations,
similarly to $age$ and \sigap, obtaining
\begin{equation}\label{eq.das.uncert.spindex}
    u(I_{\mathrm age})\,[\AA] = (0.18\pm0.11)^2 + \left( \frac{\mathrm{SNR} \; \mathrm{[\AA^{-1}]}}{15.2\pm0.6}\right)^{-(1.04\pm0.05)}
\end{equation}
(solid golden line in Fig.~\ref{f.das.uncert.c}).
Similarly to \sigap and $age$, using the stricter cut at 10~per\ cent, we would get the
same uncertainties: $0.21\pm0.12$, $14.0\pm0.6$ and $-1.09\pm0.06$.
(blue diamonds and dashed blue line in Fig.~\ref{f.das.uncert.c}).
From here on, the uncertainties on
\spindex are inferred from Eq.~(\ref{eq.das.uncert.spindex}).
A comparison between \spindex and the H\textbeta\ and $age$ is provided in Fig.~\ref{f.das.emlicom} (panels~\subref{f.das.emlicom.a} and~\subref{f.das.emlicom.b}, respectively).

\subsection{Photometric and ancillary measurements}\label{s.das.ss.phot}

Our mass and size measurements are the same as \citetalias{deugenio+2021}. We
use best-fit total magnitudes and circularised half-light radii from
Multi-Gaussian Expansion models \citep{emsellem+1994}, optimised using the
publicly available algorithm {\sc mgefit} \citep{cappellari2002}. For more
information about these measurements we refer the reader to
\citetalias{deugenio+2021}. Here we highlight three key aspects. First, these
measurements result in different FP coefficients compared to e.g. S{\'e}rsic
photometry \citepalias[][their table~3]{deugenio+2021}; this means that a
quantitative comparison
between our results and the literature is unwarranted. Second, we do not apply
a k correction; as found by \citetalias{deugenio+2021}, applying a k correction
does not reduce the FP scatter, nor alters the correlation between the FP
residuals and galaxy observables. Nevertheless, we remark that applying a k
correction does not change our conclusions \citetext{we tested both the
{\sc kcorrect} software of \citealp{blanton+roweis2007} and the polynomial
approximation of \citealp{chilingarian+2010}, which gave median k correction values of 0.048 and 0.051~mag, respectively.}.
Third, and perhaps most important point, we replace the average measurement
uncertainties from \citetalias{deugenio+2021} with the average measurement
uncertainties from \citetalias{howlett+2022}; this does not change the observed FP
scatter, but affects the relative contribution of measurement uncertainties and
intrinsic scatter.

We use four ancillary measurements. Redshift $z$ and stellar mass $M_\star$ are
taken from the SAMI DR3 catalogue \citetext{\citealp{croom+2021a}; $M_\star$ is
obtained from $i$-band absolute magnitude $M_i$ and $g-i$ colour, following
\citealp{taylor+2011} and \citealp{bryant+2015}}. The surface mass density
$\Sigma_\star$ is calculated as $M_\star / (2 \text{\textpi} R_\mathrm{e}^2)$.
Environment is measured using
\sigenv, the surface density of galaxies inside the circle enclosing the five
nearest neighbours \citep[subject to $M_r<-18.5 \, \mathrm{mag}$ within a range
of velocity $\Delta \, v < 500 \, \mathrm{km\,s^{-1}}$,][]{brough+2017}.
Note that, in \S~\ref{s.rs}, we use SDSS data and a different environment
measurement, group richness \ngroup.
Finally, we measure EW(\Halpha), the equivalent width of \Halpha, using
{\sc ppxf} to simultaneously model the continuum and emission-line spectra of
the SAMI galaxies.

\subsection{Sample selection}\label{s.das.ss.sample}

Our sample is selected as similarly as possible to that of \citet{said+2020} and
\citetalias{howlett+2022}. We
require $r$-band apparent magnitude $10 \leq m_r \leq 17$, redshift $0.0033
\leq z \leq 0.1$ (where for cluster galaxies we used the redshift of the cluster), $r$-band
S{\'e}rsic index $n\geq1.2$ (this is equivalent to requiring a concentration
$R_{90}/R_\mathrm{e} \geq 2.5$; note that not all SAMI galaxies have $i$-band S{\'e}rsic 
photometry, so we drop the similar requirement in this band), axis ratio $b/a\geq0.3$
in both $r$- and $i$-band\footnote{A cut in galaxy shape is standard practice in FP
cosmology, to avoid intrinsically flattened systems seen close to edge-on \citetext{
\citealp{said+2020}, \citetalias{howlett+2022}}},
colour $g-i\geq0.73 - 0.02(M_r + 20)$, and aperture
velocity dispersion $\sigap\geq70\,\mathrm{km\,s^{-1}}$. We remove
galaxies with strong \Halpha emission lines by requiring that the
equivalent width is EW(\Halpha)$>$-1~\AA. We further
require that our galaxies are classified as early-type in their optical morphology,
i.e.\ $0 \leq \mathrm{mtype} \leq 1$ \citep[in the SAMI morphological 
classification, ellipticals have mtype=0 and lenticulars have mtype=1; 
see][]{cortese+2016}.
These selection criteria result in 703~galaxies for each of the nine mock datasets
with exposure times 0.125, 0.1875, \dots, 2~hours; the mock sample with
$\texp=3$~hours contains only 665~galaxies, because 38~galaxies had SAMI
observations with exposure time $\texp[SAMI]<3$~hours.

\subsection{Modelling the 3- and 4-dimensional planes}\label{s.das.ss.methods}

Systematics due to model choice dominate the uncertainties in the parameters of the 
FP \citepalias{deugenio+2021}. The three most popular models are
an infinite plane with Gaussian scatter
\citetext{either orthogonal or along the z axis,
\citealp{cappellari+2013a}, \citealp{said+2020}, \citetalias{deugenio+2021};
this model underlies $\chi^2$ minimisation algorithms},
an infinite plane with Laplacian scatter \citep[usually orthogonal to the plane,][]{
degraaff+2020, degraaff+2021}, and a 3-d Gaussian \citetext{with or without
censoring, \citealp{saglia+2001}, \citealp{magoulas+2012}, \citealp{said+2020},
\citetalias{deugenio+2021}}. However, despite their differences, these models
have in common that the residuals about the best-fit FP correlate most strongly
with stellar population age \citepalias{deugenio+2021}. For this reason, in this
work we use only a single model, based on the least trimmed squares, a robust
$\chi^2$ minimisation algorithm \citetext{LTS, \citealp{rousseeuw+driessen2006};
we use the free implementation {\sc \href{https://pypi.org/project/ltsfit/}{
lts\_planefit}}, \citealp{cappellari+2013a}}. The plane equation for the model
underlying this `direct-fit' method can be written as
\begin{equation}\label{eq.das.3dplane}
    r = a\,s + b\,i + d
\end{equation}
where, from now on, $s$, $i$ and $r$ are the logarithms of \sigap, $\langle I
\rangle_\mathrm{e}$ and $R_\mathrm{e}$. In the context of this work, this
algorithm presents a key advantage over a multivariate Gaussian model: while
extending a 3-d Gaussian to four dimensions increases the number of
parameters from six to ten, extending Eq.~(\ref{eq.das.3dplane}) to four
dimensions only requires one more parameter
\begin{equation}\label{eq.das.4dplane}
    r = a\,s + b\,i + c\,A + d
\end{equation}
where $A$ denotes the logarithm of $age$. We refer to this model as the
4-d hyperplane (dropping the adjective `fundamental'). We also consider the
hyperplanes where $A$ is replaced by \spindex, or by $\upsilon_\star \equiv
\log \, \Upsilon_\star$.
We infer the uncertainties on the plane and hyperplane parameters from the
measurement uncertainties on the data, which we assume to be Gaussian (we call these the `formal uncertainties'). These
numbers, labelled $u(a), u(b), \dots$ are reported as one standard deviation
in the figures.
For the mocks with 1~h exposure, we validated these uncertainties against
bootstrapping the sample one hundred times. The two methods are in good
agreement, therefore we adopted everywhere the formal uncertainties, which are
less computationally demanding.

\section{The hyperplane of early-type galaxies}\label{s.rh}

To determine the impact of data quality on the 3-d FP and on the 4-d hyperplane,
we find the best 3-d and 4-d fit for each of the ten sets of mock observations.
For brevity, we show only example fits for the mock with 1~hour exposure time
(\S~\ref{s.rh.ss.planes}). We then study the impact of spectral $SNR$ on the FP
and hyperplane coefficients (\S~\ref{s.rh.ss.fits}) and, most importantly, on the
observed scatter (\S~\ref{s.rh.ss.rms}).
We then study the presence and
significance of correlations between the fit residuals and two key galaxy
observables: stellar-population age and local environment density
(\S~\ref{s.re}; once again, in the interest of brevity, we only show
the results for the mock with $\texp = 1$~hour).
Finally, we study the effect of these two variables on the FP residuals, to
assess whether they are independent (\S~\ref{s.re.ss.envage}).

\subsection{Example best-fit planes}\label{s.rh.ss.planes}

\begin{figure*}
    \centering
    \includegraphics[trim={0 0 0 1cm},clip,width=\textwidth]{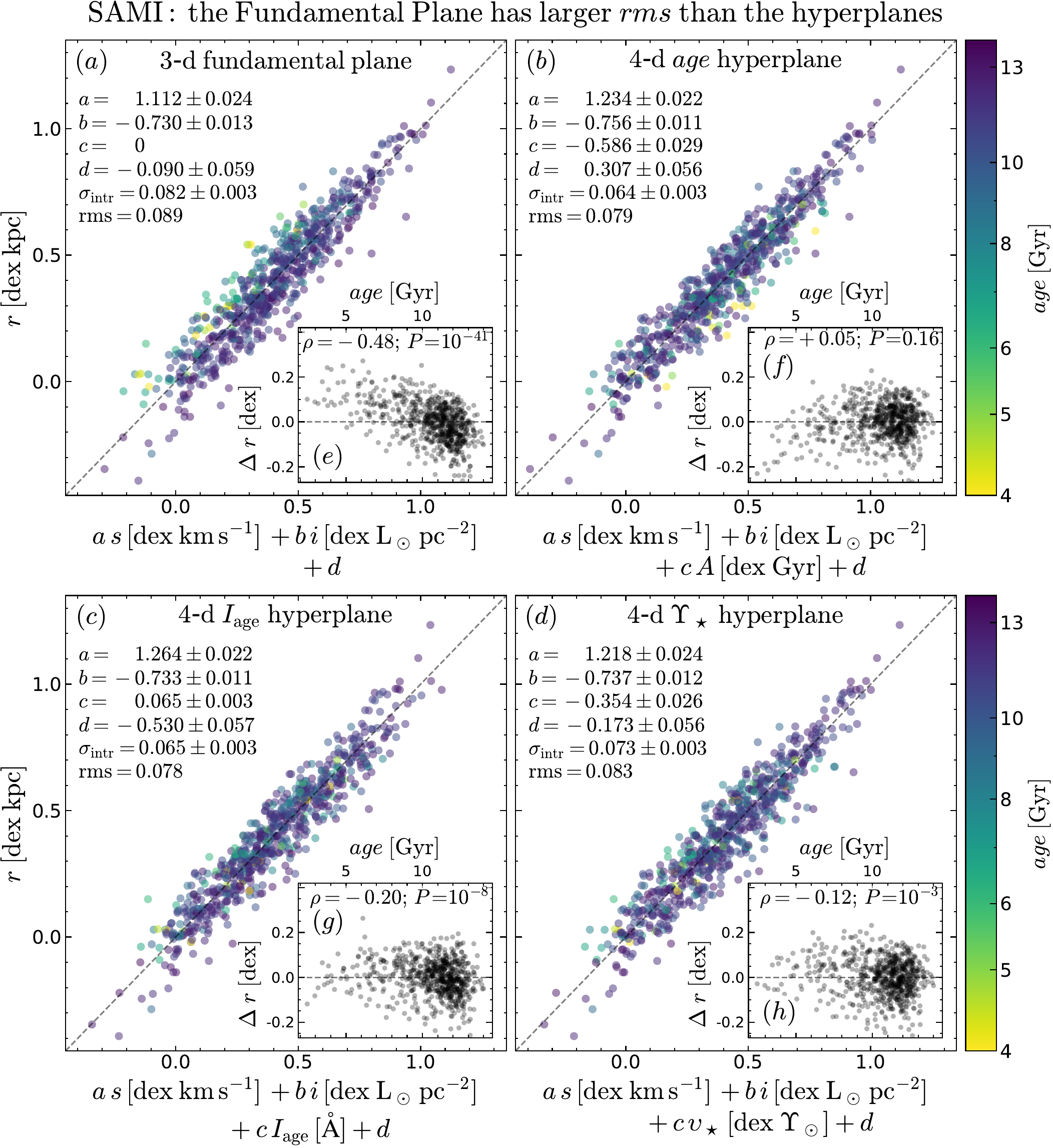}
    {\phantomsubcaption\label{f.rh.plane.a}
     \phantomsubcaption\label{f.rh.plane.b}
     \phantomsubcaption\label{f.rh.plane.c}
     \phantomsubcaption\label{f.rh.plane.d}
     \phantomsubcaption\label{f.rh.plane.e}
     \phantomsubcaption\label{f.rh.plane.f}
     \phantomsubcaption\label{f.rh.plane.g}
     \phantomsubcaption\label{f.rh.plane.h}}
    \caption{Predicted vs measured galaxy size, comparing the best-fit
    3-d FP (panel~\subref{f.rh.plane.a}) to the 4-d hyperplane using either
    $age$ (panel~\subref{f.rh.plane.b}), \spindex
    (panel~\subref{f.rh.plane.c}), or $\Upsilon_\star$ (panel~\subref{f.rh.plane.d}).
    Using fibres with 1.5-arcsec radius and an integration
    time of 1~hour with AAT/SAMI, the hyperplanes have both lower intrinsic scatter
    \sigintr and lower observed $rms$. Galaxies are colour-coded
    according to light-weighted stellar-population age. The inset
    panels~\subref{f.rh.plane.e}--\subref{f.rh.plane.h}
    show the residuals \Dr as a function of $age$: the correlation with
    the largest-magnitude coefficient and highest significance is for the FP.
    For \spindex and $\Upsilon_\star$, the correlation is weaker and for
    the $age$ hyperplane there is no correlation (see\ Fig.~\ref{f.rh.cores.b}).
    }\label{f.rh.plane}
\end{figure*}

In Fig.~\ref{f.rh.plane} we show the FP and the three hyperplanes for the mock
with $\texp = 1$~hour. Each panel shows measured
vs predicted $r$, for the 3-d FP (panel~\subref{f.rh.plane.a}), for the 4-d
$age$ hyperplane (panel~\subref{f.rh.plane.b}), for the 4-d hyperplane using
\spindex (panel~\subref{f.rh.plane.c}), and for the 4-d hyperplane using
mass-to-light ratio (panel~\subref{f.rh.plane.d}).
Each of the 703 circles represents one SAMI galaxy, colour-coded by $age$. The
best-fit coefficients
from Eq.s~(\ref{eq.das.3dplane}--\ref{eq.das.4dplane}) are reported in the top
left corner of each panel (for the FP, $c=0$ by construction).
We find $\sigma_\mathrm{intr} = 0.082\pm0.003$, $0.064\pm0.003$, $0.065\pm0.003$
and $0.073\pm0.003$ respectively for the FP and for the age, index and
mass-to-light ratio hyperplanes. This demonstrates that, for the adopted
observing setup (1-hour integration on a 4-metre telescope), the 4-d hyperplane
is able to utilise stellar-population information to reduce \sigintr. The
magnitude of this improvement depends directly on our assessment of the random
uncertainties, which are notoriously difficult to estimate accurately:
overestimating the uncertainties would underestimate \sigintr. But this does not
apply to the \textit{observed} scatter (labelled $rms$ in Fig.~\ref{f.rh.plane}).
We find $rms = 0.089$, $0.079$, $0.078$ and $0.083$, so the 4-d hyperplanes are 
tighter than the FP --- even accounting for the measurement uncertainties on
$A$, \spindex and $\upsilon_\star$. This reduced $rms$ suggests that the
hyperplanes could be used to derive more precise distances than the FP
(\S~\ref{s.d.ss.distun}).

The inset panels~\subref{f.rh.plane.e}--\subref{f.rh.plane.h} show the fit
residuals ($\Dr \equiv r - r_\mathrm{predicted}$) as a function of $age$.
For the FP there is a statistically significant correlation between \Dr 
and $age$ (panel~\subref{f.rh.plane.e}), which can also be appreciated as the
clear trend in colour hue across the FP (panel~\subref{f.rh.plane.a}). This
suggests that $age$ contains information useful to reduce the FP scatter.
In contrast, for the residuals of the hyperplanes, correlations with $age$
are weaker (panel~\subref{f.rh.plane.g} and \subref{f.rh.plane.f}) or not
significant (panel~\subref{f.rh.plane.f}). This fact, coupled with the
observation that the hyperplanes have smaller $rms$ than the FP, confirms that
the hyperplanes are tighter owing to the inclusion of stellar population
information.

One may argue whether this stellar-population information is simply $M_\star$,
given the widely reported reduction in scatter of the stellar-mass plane
\citep{hyde+bernardi2009},
obtained by replacing $i$ with $\log \, \Sigma_\star$.
For our sample, the stellar-mass plane shows similar $rms$ as the FP, but smaller \sigintr.
A one-to-one comparison between the standard FP and the stellar-mass plane is
reported in Appendix~\ref{app.mstarplane}.

\subsection{Best-fit parameters as a function of exposure time}\label{s.rh.ss.fits}

\begin{figure}
    \centering
    \includegraphics[trim={0 0 0 1cm},clip,width=\columnwidth]{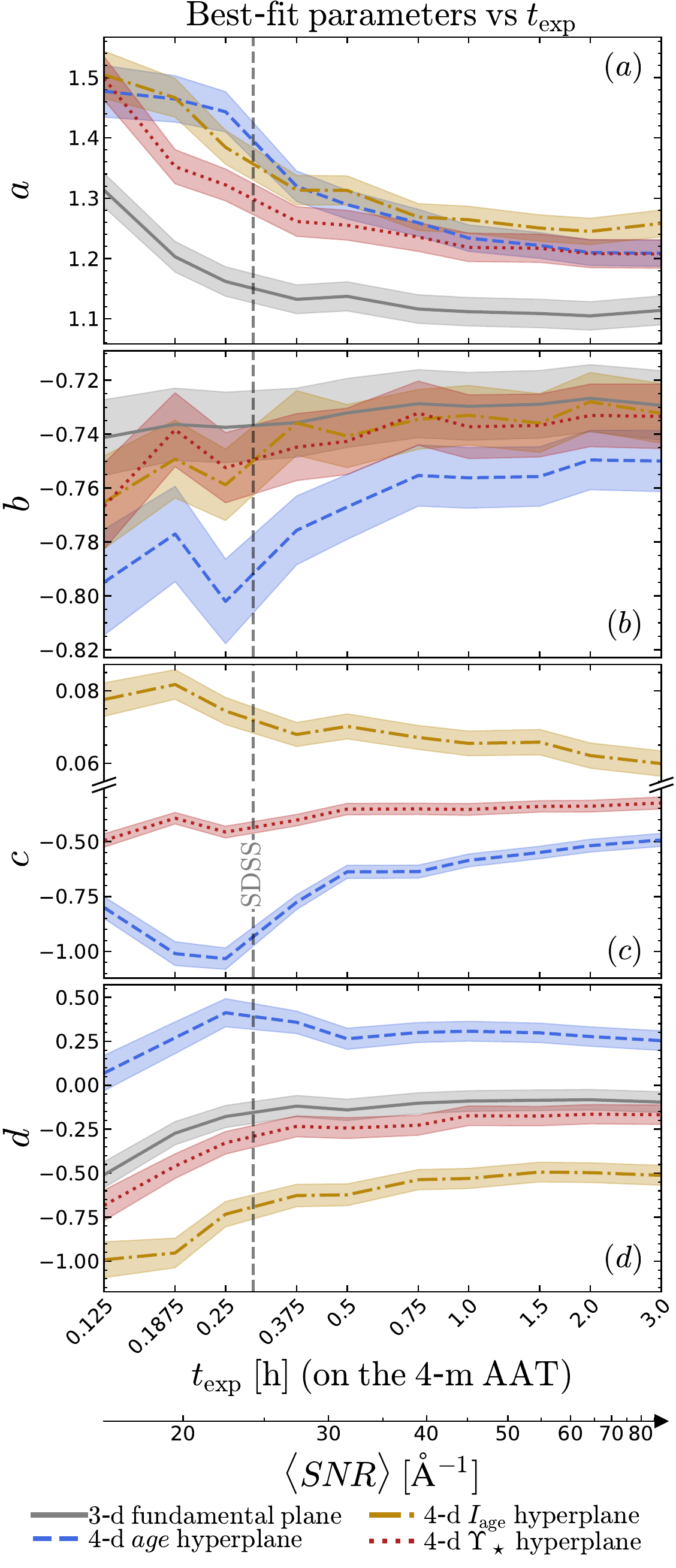}
    {\phantomsubcaption\label{f.rh.fits.a}
     \phantomsubcaption\label{f.rh.fits.b}
     \phantomsubcaption\label{f.rh.fits.c}
     \phantomsubcaption\label{f.rh.fits.d}
    }
    \caption{The best-fit FP and hyperplane coefficients as a function of the
    mock exposure time \texp (and of the average sample $SNR$, $\langle
    SNR \rangle$). The dashed vertical line is \texp[SDSS]. For
    $\texp < \texp[SDSS]$, the best-fit coefficients can be
    significantly different from their limiting values at high $SNR$. For the
    FP, convergence is already achieved at $\texp = \texp[SDSS]$;
    for the hyperplanes, convergence requires $\texp \gtrsim
    0.5$~hours. Note the broken y axis in panel~\subref{f.rh.fits.c}, to
    accommodate two different scales.
    }\label{f.rh.fits}
\end{figure}

In Fig.~\ref{f.rh.fits} we show the best-fit coefficients as a function of \texp
(and, equivalently, as a function of the median $SNR$ over the whole sample,
$\langle SNR \rangle$, reported on the bottom axis). As a benchmark, we report
the equivalent exposure time of SDSS spectroscopy (vertical dashed line, see
\S~\ref{s.das.ss.snr}).
Each of panels~\subref{f.rh.fits.a}--\subref{f.rh.fits.d} traces the homonym plane
coefficient (cf. Eq.s~(\ref{eq.das.3dplane}--\ref{eq.das.4dplane}); for the FP, 
$c=0$ and is not shown). The solid black line is for the FP, with the gray
shaded regions encompassing the statistical uncertainties. We note that the FP
coefficients depend on the quality of the spectra, particularly below the median
$SNR$ of SDSS. To find the value of $a$ for
\texp[SDSS], we interpolate linearly between $\texp = 0.25$ and
0.375~hours and get $a = 1.142$. This is 2.5~per cent larger than the limiting
value at $\texp=3$~hours (we find the same using the mean of all mocks
with \texp$>0.375$~hours). To our knowledge, this is the first time this effect 
is reported in the literature.\footnote{This bias goes in the same direction as
the reported difference in $a$ between the optical and NIR FPs
\citepalias{magoulas+2012}, because the (optical) spectroscopy of the NIR
FP has lower $SNR$ than the (optical) spectroscopy of the optical FP, and the
NIR-derived $a$ is larger than the optical-derived $a$. Quantifying how much
the effect we report contributes to the difference between the optical and NIR
FPs is beyond the scope of this work.}
We find similar effects for $b$ and $d$.
The best-fit coefficients for the hyperplanes also show similar effects. We
show the $age$ hyperplane (dashed blue lines in Fig.~\ref{f.rh.fits}), the
\spindex hyperplane (dash-dotted golden line) and the $\upsilon_\star$
hyperplane (dotted red lines). Compared to the FP coefficients, the hyperplanes
coefficients converge at longer \texp, in the range $0.5 < \texp
< 1$~hour, depending on the coefficient and hyperplane flavour.
The non-monotonic behaviour of $b$, $c$ and $d$ below \texp[SDSS] is
due to the hyperplanes not converging for samples with $\langle SNR \rangle <
25$~\AA$^{-1}$.
Overall, we find that the FP parameters are very close to their `asymptotic' values already
for $\langle SNR \rangle \approx 25~\AA^{-1}$, even though percent-level systematics seem to persist until a much higher $\langle SNR \rangle \approx 40~\AA^{-1}$.
It may be worth investigating whether a Bayesian approach can model out this bias, for instance, by introducing priors based on high-SNR observations.

\subsection{Observed scatter and residual correlations as a function of exposure time}\label{s.rh.ss.rms}

\begin{figure}
    \centering
    \includegraphics[trim={0 0 0 0.9cm},clip,width=\columnwidth]{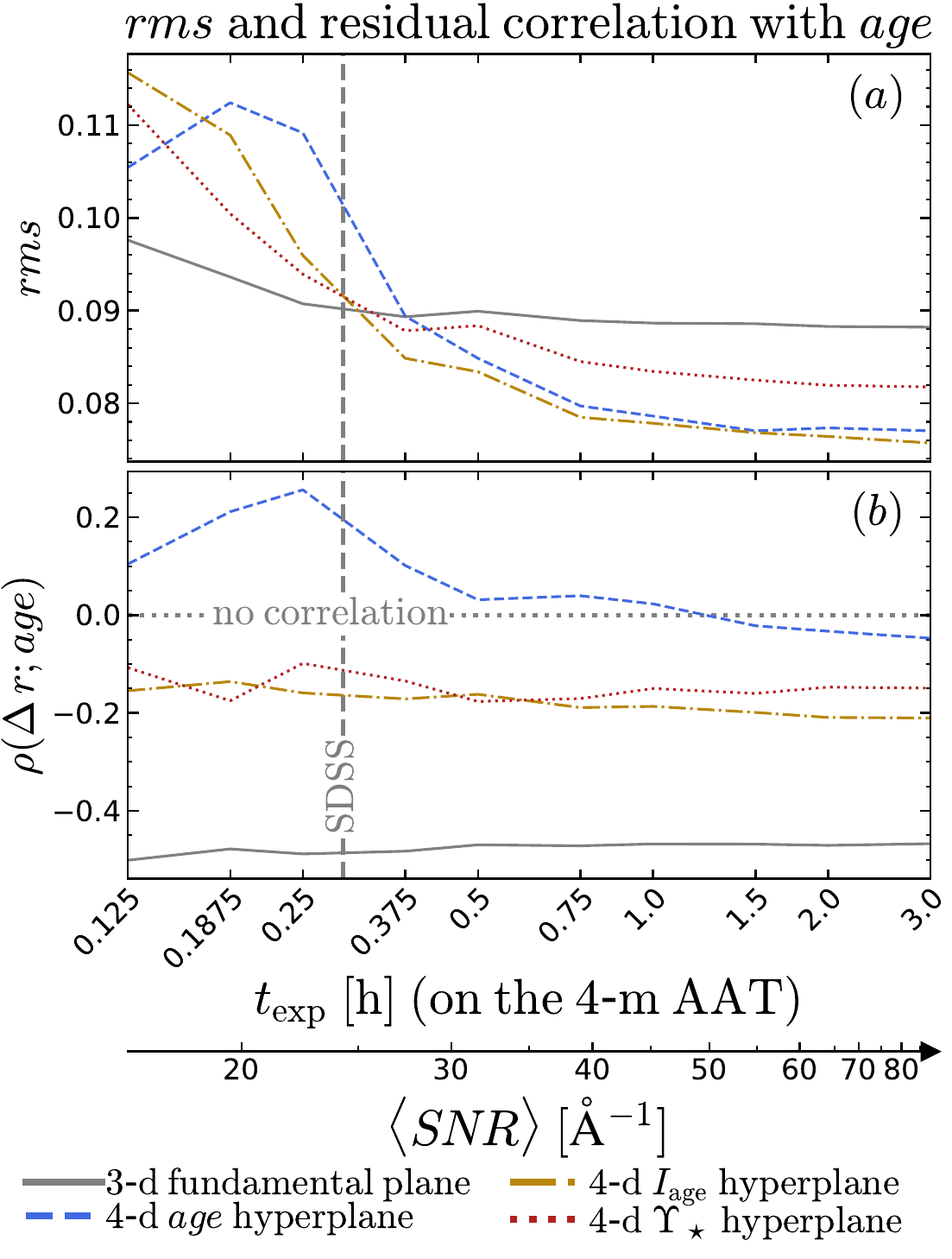}
    {\phantomsubcaption\label{f.rh.cores.a}
     \phantomsubcaption\label{f.rh.cores.b}
    }
    \caption{$rms$ (panel~\subref{f.rh.cores.a}) and residuals--$age$
    correlation coefficient ($\rho(\Dr;\,age)$;
    panel~\subref{f.rh.cores.b}) for the best-fit FP and hyperplanes, as a
    function of \texp (and $\langle SNR \rangle$). The lines are the same as
    in Fig.~\ref{f.rh.fits}.
    For $\texp < \texp[SDSS]$, the FP (solid black line) has
    smaller $rms$ than any of the hyperplanes (panel~\subref{f.rh.cores.a}). For
    $\texp \sim \texp[SDSS]$, the \spindex and $\Upsilon_\star$
    hyperplanes have similar $rms$ to the FP (dashed-dotted golden line and
    dotted red line). Finally, for $\texp > \texp[SDSS]$, all the
    hyperplanes (including the $age$ hyperplane, dashed blue line) have smaller
    $rms$ than the FP. Regardless of \texp, the FP always displays a correlation
    between the residuals \Dr and $age$ (panel~\subref{f.rh.cores.b}).
    Adding a fourth observable to the FP reduces the magnitude of this
    correlation, even with short \texp (low $\langle SNR \rangle$). However, to
    also reduce the $rms$, this fourth observable must be sufficiently precisely
    determined that its random measurement uncertainties do not negatively
    offset the gain due to the additional information they provide.
    }\label{f.rh.cores}
\end{figure}

In Fig.~\ref{f.rh.cores} we study the $rms$ and residual correlations with $age$
as a function of \texp. From panel~\subref{f.rh.cores.a}, we see that the FP
$rms$ decreases only by $\approx$10~per cent across the whole range in \texp;
in particular, the FP $rms$ converges to its minimum $rms$ already near
\texp[SDSS] (to within 2~per cent). The hyperplanes show different
behaviours: they have larger/smaller $rms$ than the FP for short/long \texp.
For the $age$ hyperplane, equality is reached at $\texp \approx
0.375$~hours ($\langle SNR \rangle \approx 28$~\AA$^{-1}$). For the other two
hyperplanes, our mocks show that equality is already reached near
\texp[SDSS] (vertical dashed line). For \spindex, we tested this
prediction using archival SDSS data and found it to be accurate
(\S~\ref{s.rs}).
Convergence to the minimum $rms$ occurs in the range
0.75--1.5~hours, i.e. for $\langle SNR \rangle \approx$ 40--55~\AA$^{-1}$.

In agreement with \citetalias{deugenio+2021}, we find that the best
predictor\footnote{Here and afterwards, the `best predictor' of \Dr has the
correlation coefficient with the largest magnitude and highest statistical
significance, as in \citetalias{deugenio+2021}.}
of the FP \Dr is $age$. Indeed, for the FP, the anticorrelation has Spearman
correlation coefficient $\rho(\Dr;age) = -0.5$--$0.45$
(panel~\subref{f.rh.cores.b}). For the $age$
hyperplane, we find positive correlation below 0.5~hours, and no correlation
above it, suggesting that age information has been used to reduce the $rms$.
Interestingly, for the other two hyperplanes, \Dr still anticorrelates
with $age$ (albeit with less than half the FP magnitude). For the
$\Upsilon_\star$ hyperplane, one can argue $age$ information was not correctly
encapsulated in the fit, which explains why its $rms$ is larger than for the
$age$ hyperplane. However, for the \spindex hyperplane, the $rms$ is the same
as the $age$ hyperplane, yet $\rho$ has larger magnitude. This means that
\spindex taps in part information other than $age$ to reduce the FP scatter.

To first order, the precision of the FP as a distance indicator can be gauged
from its observed $rms$. Comparing this value between the three hyperplanes
(Fig.~\subref{f.rh.cores.a}), we find that $\Upsilon_\star$ yields the largest
$rms$ of the three.
The other two implementations, the $age$ and \spindex hyperplanes are comparable
to one another and have substantially lower $rms$, but \spindex has
the critical advantage of being an empirical observable, measurable directly
from the spectra without any additional assumptions (cf.\ 
\S~\ref{s.das.ss.hfm.ext}).
For this reason, from now on, we focus on the \spindex hyperplane.

\section{Residual correlations with environment}\label{s.re}

The existence of residual correlations between \Dr and local environment
\citetext{\citetalias{magoulas+2012}, \citetalias{howlett+2022}} is worrying for
cosmological applications, because FP-derived distances would place galaxies in
under/overdense environments systematically closer/further than their true
distance. In this section, we study the strength and significance of this
residual correlation for the FP and for the \spindex hyperplane. As a
comparison, we use the \Dr--$age$ anticorrelation, the best predictor of \Dr.
As a measure of local environment, we use \sigenv (\S~\ref{s.das.ss.phot}).

\begin{figure}
    \centering
    \includegraphics[width=\columnwidth]{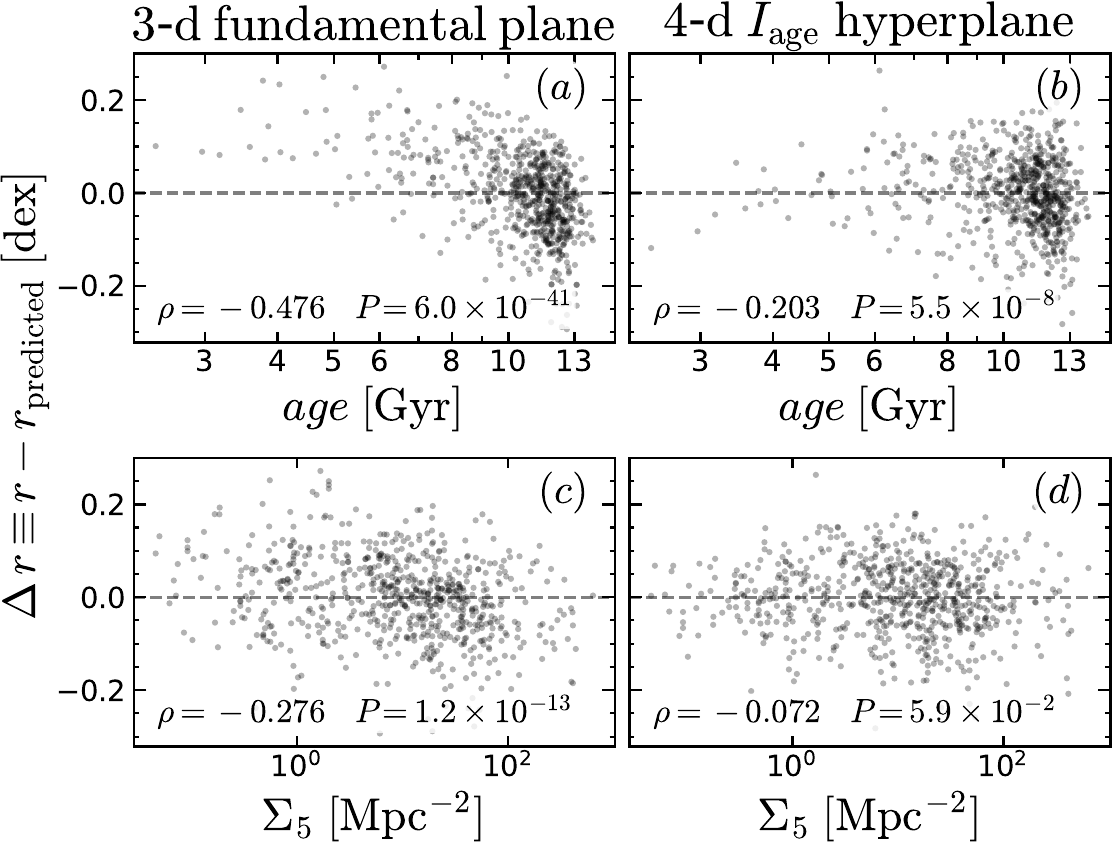}
    {\phantomsubcaption\label{f.re.resid.a}
     \phantomsubcaption\label{f.re.resid.b}
     \phantomsubcaption\label{f.re.resid.c}
     \phantomsubcaption\label{f.re.resid.d}}
    \caption{
    Residuals \Dr of the 3-d FP (left column) and of the 4-d \spindex hyperplane
    (right column) as a function of stellar population age (top row)
    and local environment density (bottom row). Each panel also reports
    the Spearman rank correlation coefficient $\rho$ and the corresponding
    $P$ value. For the FP, \Dr correlates with both $age$ and \sigenv. For the
    hyperplane, \Dr has a much weaker (but still significant) correlation with
    $age$, while the correlation with \sigenv is weaker and only marginally
    significant. Note panels~\subref{f.re.resid.a} and~\subref{f.re.resid.b} are
    the same as panels~\subref{f.rh.plane.e} and~\subref{f.rh.plane.g} from
    Fig.~\ref{f.rh.plane}, and are reproduced here for convenience.
    }
    \label{f.re.resid}
\end{figure}

In the left column of Fig.~\ref{f.re.resid} we show the FP \Dr as a function of
$age$ (panel~\subref{f.re.resid.a}) and \sigenv (panel~\subref{f.re.resid.c}).
Each panel reports the value of $\rho$ and its
significance $P$. Both $age$ and \sigenv correlate with \Dr, but, in
agreement with \citetalias{deugenio+2021}, the correlation with $age$ has
larger magnitude and higher statistical significance.

In the right column of Fig.~\ref{f.re.resid}, we show \Dr for the \spindex
hyperplane. For $age$, $\rho$ is now less than half the FP value (and its
statistical significance is also lower, panel~\subref{f.re.resid.b}); for
\sigenv, the change is even more dramatic: $\rho$ is only about one fourth
the FP value, and its significance is marginal (2-\textsigma,
panel~\subref{f.re.resid.d}).

This result shows that the correlations of \Dr with $age$ and \sigenv are not
independent. By correcting the FP for \spindex (which encodes in part $age$
information) the 4-d hyperplane also greatly reduces, or even eliminates, the
problem of the environment bias of FP distances.

\subsection{Relation between age and local environment}\label{s.re.ss.envage}

Given the correlation between environment density and $age$, which is the
physical driver between the \Dr--\sigenv and \Dr--$age$ correlations? To
address this question, we use the method of partial correlation coefficients
\citep[hereafter, PCC; see e.g.][]{bait+2017, bluck+2019, baker+2022}.
In general, if two variables $x$ and $z$ are both independently correlated with
a third variable $y$, then this will induce an 
apparent correlation between $y$ and $z$. PCCs address this issue by quantifying
the strength and significance of the correlation between $y$ and $z$ while
controlling for $x$. Similarly to the standard Spearman rank correlation
coefficient, a value of zero implies no correlation, and -/+ 1 implies perfect
anti/correlation. In the following, we denote with $\rho(x,z \vert y)$ the
partial correlation coefficient between $x$ and $z$ removing the effect of $y$.

\begin{figure}
    \centering
    \includegraphics[width=\columnwidth]{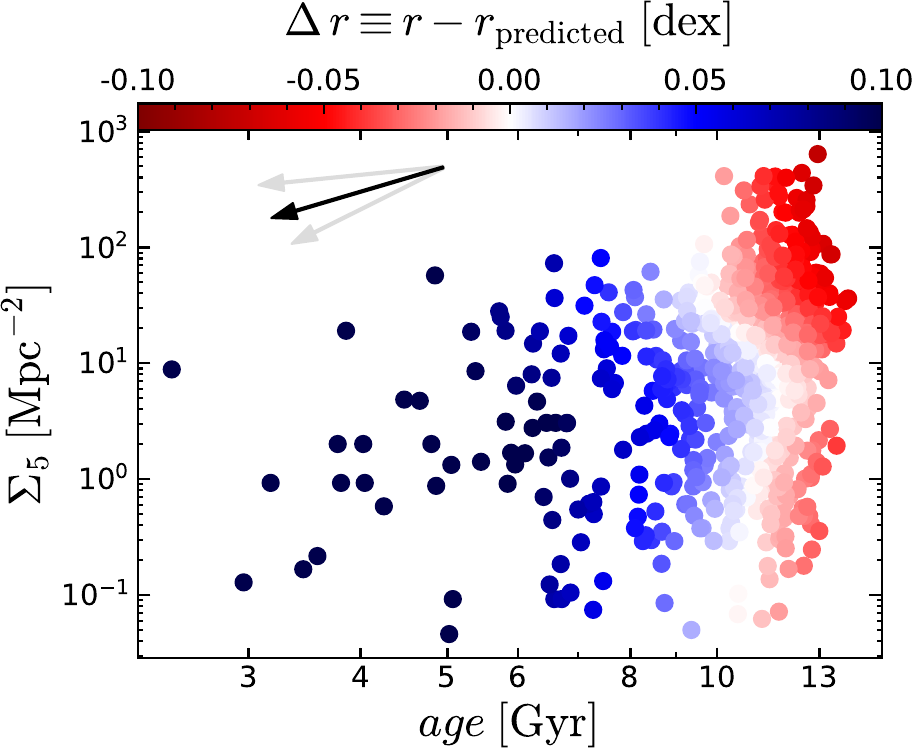}
    \caption{Distribution of the FP residuals \Dr on the $age$--\sigenv
    plane. \Dr has been LOESS smoothed to highlight the trend. The arrow
    is a graphical representation of the partial correlation coefficients;
    it has an angle of $(197\pm 12)^\circ$, with the uncertainties
    encompassing the 1\textsuperscript{st} and 99\textsuperscript{th}
    percentiles. An angle of $180^\circ$ would correspond to perfect
    anti-correlation between \Dr and $age$, and no independent correlation
    with \sigenv. \Dr is strongly dominated by $age$, with only a weak
    independent anti-correlation with \sigenv.
    }\label{f.re.parcorr}
\end{figure}

In Fig.~\ref{f.re.parcorr}, we show the $age$--\sigenv plane, colour-coded by
the FP residuals \Dr \citetext{here \Dr has been smoothed using the robust
locally-weighted regression algorithm of \citealp{cleveland+devlin1988} and
\citealp{cappellari+2013b}, LOESS}. Regions of uniform colour hue are tilted
with respect to the axes, highlighting that $age$ and \sigenv have independent
roles. Controlling for \sigenv, we find
$\rho(\Dr,age \vert \sigenv) = -0.44$, with $P=9.9\times 10^{-34}$,
whereas controlling for $age$, we find $\rho(\Dr,\sigenv \vert age) =
-0.14$, with $P=3.4\times 10^{-4}$. Thus \Dr correlates independently with both
$age$ and \sigenv, but the correlation with $age$ is both stronger
(higher-magnitude correlation coefficient) and more significant. The relative
importance of the two correlations can be visualised using the arrow
representation of the ratio of PCCs \citep[introduced by][]{bluck+2020a}. We
obtain an angle of $(197\pm12)^\circ$ (black arrow in Fig.~\ref{f.re.parcorr};
the grey arrows represent the 1\textsuperscript{st}--99\textsuperscript{th}
percentiles from bootstrapping the sample one thousand times). The arrow is
almost parallel to the $age$ axis, indicating that for the FP \Dr is driven
almost entirely by $age$, and the independent correlation with environment plays
only a secondary role, if at all.
Thus the \Dr--\sigenv anticorrelation is mostly due to the known correlation
between the \Dr and $age$, combined with the correlation between $age$ and
\sigenv. This explains why, in correcting for the $age$ bias, the 4-d
hyperplanes provide the environment bias correction for  free
(\S~\ref{s.re}).

\section{Application to the SDSS sample}\label{s.rs}

In the previous two sections we have made two key statements:
\begin{itemize}
  \item The hyperplanes have lower $rms$ than the FP (for sufficiently high
        $SNR$; \S~\ref{s.rh.ss.rms}, Fig.s~\ref{f.rh.plane} and~\ref{f.rh.cores});
  \item By removing the $age$ bias from the FP, the hyperplanes also remove the
	environment bias (as measured by \sigenv; \S~\ref{s.re},
        Fig.~\ref{f.re.resid})
\end{itemize}
Both these statements are based on data from SAMI, which we degraded to simulate
the effect of lower-$SNR$, single-fibre observations. In this section, we test
these two hypotheses using independent data from SDSS. The sample consists of
34,059 galaxies from \citetalias{howlett+2022}. This sample is originally drawn from
the SDSS Data Release 14 \citep{abolfathi+2018}, with the criteria very similar 
to those summarised in \S~\ref{s.das.ss.sample}. To these measurements, we added
our own Lick-index and $age$ measurements, using the methods outlined in
\S~\ref{s.das.ss.hfm.ext}. This results in a sample of 31,909 galaxies,
There is a key difference between this sample and SAMI; for SAMI, we measure
\emph{local} environment (using \sigenv), but for SDSS we measure \emph{global}
environment \citep[using \ngroup, the number of galaxies in the
group,][]{tempel+2017}.

\begin{figure*}
    \centering
    \includegraphics[trim={0 0 0 1.1cm},clip,width=\textwidth]{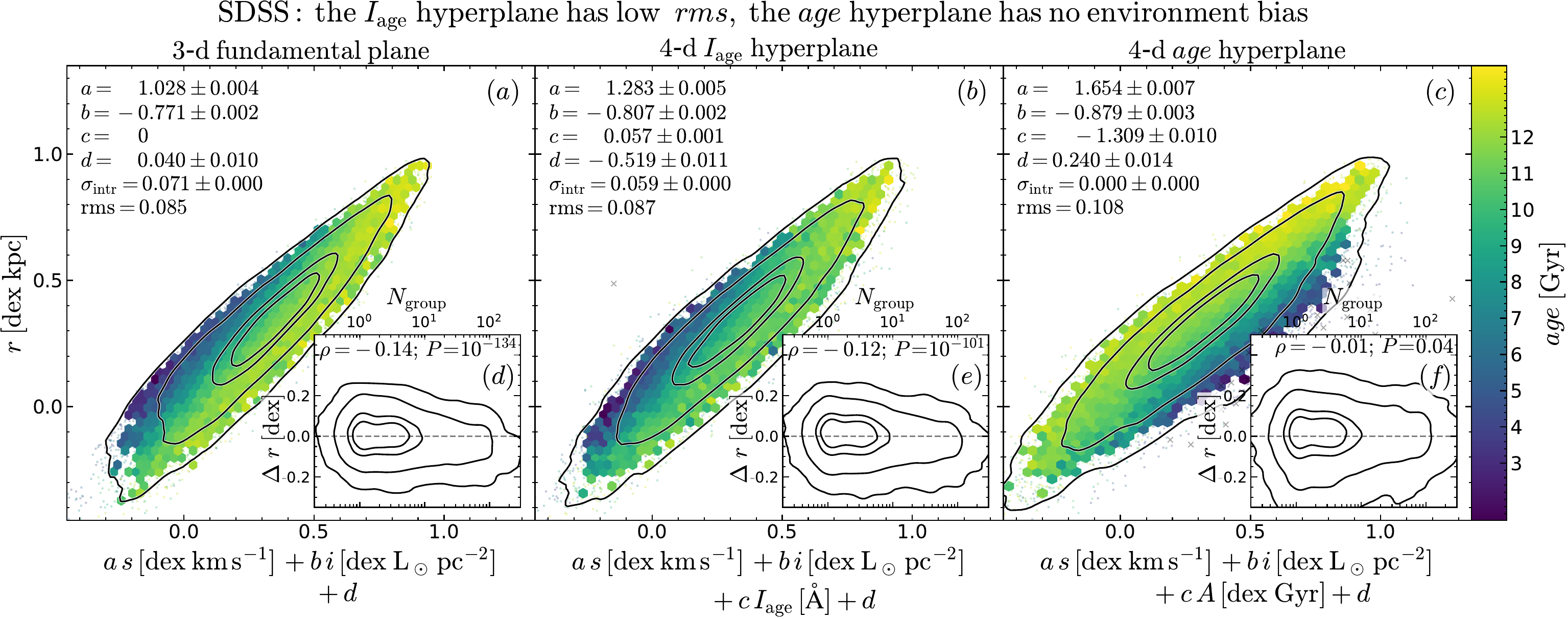}
    {\phantomsubcaption\label{f.rs.plane.a}
     \phantomsubcaption\label{f.rs.plane.b}
     \phantomsubcaption\label{f.rs.plane.c}
     \phantomsubcaption\label{f.rs.plane.d}
     \phantomsubcaption\label{f.rs.plane.e}
     \phantomsubcaption\label{f.rs.plane.f}}
    \caption{
    Predicted vs measured galaxy size using data from SDSS. The 3-d FP
    (panel~\subref{f.rs.plane.a}) has the same $rms$ as the 4-d \spindex
    hyperplane (panel~\subref{f.rs.plane.b}). The median uncertainty of the 
    two samples is comparable, in agreement with the results of our mock
    observations (see \S~\ref{s.d.ss.distun}). In contrast, the $age$
    hyperplane has larger $rms$, also in agreement with SAMI-based predictions.
    Bins contain a minimum of five
    galaxies and are colour coded by their mean $age$; individual galaxies
    are represented by dots. The inset panels show the correlation between
    the fit residuals and global environment \ngroup. The FP and
    \spindex hyperplanes display the correlation reported by
    \citetalias{howlett+2022}, whereas for the $age$ hyperplane there is no
    residual correlation.
    }\label{f.rs.plane}
\end{figure*}

In \S~\ref{s.rh.ss.rms}, we have shown that the \spindex hyperplane has the same
$rms$ of the FP, already for exposure time $\texp = 0.3$~hours on the 4-metre
AAT Telescope. This is roughly equivalent to the depth of the SDSS data
(0.75~hours on the 2.5-metre APO Telescope; dashed vertical line in
Fig.~\ref{f.rh.cores.a}). The FP and the \spindex hyperplane are shown in
Fig.~\ref{f.rs.plane}; as predicted from the $rms$--\texp relation for
\texp=\texp[SDSS] (Fig.~\ref{f.rh.cores.b}), these two scaling relations
have comparable $rms$: 0.085~dex for
the FP (panel~\subref{f.rs.plane.a}) and 0.087~dex for the hyperplane
(panel~\subref{f.rs.plane.b}); the median distance uncertainties is 0.085~dex
for both distance indicators.

At the large-size (high-mass) end of the FP, adding \spindex information
\textit{increases} the scatter. This is what we expect if the scatter is
driven by stellar mass-to-light ratio, which is highly degenerate at the 
high-mass end and therefore requires high-SNR measurements to disentangle.
Conversely, at the small-size (low-mass) end of the FP, adding \spindex
information reduces the $rms$. Overall, at the SNR level of the 
SDSS data, the two effects compensate and the hyperplane has comparable $rms$
to the FP. 
We further notice that in the FP, the oldest galaxies (yellow hues in
Fig.~\ref{f.rs.plane}) lie preferentially below the plane, forming a tighter
sequence with steeper slope; conversely, the youngest galaxies (blue hues)
lie preferentially above the plane. So these two extremes of the $age$
distribution occupy different regions of the parameter space, such that
selecting on observables that correlate with $age$ (e.g., \sigap) may 
alter the best-fit FP slope \citep{donofrio+2008}.

Crucially, when we study the fit residuals, we find a statistically significant
trend with environment for \emph{both} the FP and the \spindex hyperplane,
unlike what we found based on SAMI data (Fig.~\ref{f.rs.plane.d}
and~\subref{f.rs.plane.e}). The magnitude of the Spearman's rank
correlation coefficient is largest for the FP residuals, so the \spindex
hyperplane reduces the environment bias. However, the remaining bias is still
large ($\rho = -0.118$) and highly significant.

\subsection{The SDSS \texorpdfstring{$age$}{\textit{age}} hyperplane}\label{s.rs.ss.sdssage}

Fig~\ref{f.rs.plane.c} shows the $age$ hyperplane for SDSS, compared again
to the FP. Here the $rms$ and distance uncertainties are larger than for the FP 
--- again in agreement with the SAMI predictions, given the depth of the SDSS
spectra (Fig.~\ref{f.rh.cores.a}).
However, for the $age$ hyperplane, the correlation between the residuals and
environment is only marginally significant ($\rho = -0.01$, $P=0.04$); comparing
these values to the corresponding results for the FP ($rho = -0.137$,
$P<10^{-10}$), we conclude that the $age$ hyperplane is effectively free from
the environment bias affecting the FP.

Using partial correlation coefficients to study the interplay between the FP
\Dr, $age$ and \ngroup, we find that
$\rho(\Dr,age \vert \ngroup) = -0.45$, whereas controlling
for $age$, we find $\rho(\Dr,\ngroup \vert age) = -0.09$. While both results
are statistically significant, this shows that the correlation of the FP
residuals \Dr with \ngroup is driven primarily by the combination of the
\Dr--$age$ correlation with the $age$--\ngroup correlation.
Note that --- in agreement with \citetalias{howlett+2022} --- we find that absolute
magnitude $M_r$ does not explain the \Dr--\ngroup correlation; we find
$\rho(\Dr,\ngroup \vert M_r) = -0.15$, very similar to the regular Spearman's
rank correlation coefficient between \Dr and \ngroup ($\rho = -0.137$,
Fig.~\ref{f.rs.plane.d}).

\section{Discussion}\label{s.d}

\subsection{Physical meaning: \texorpdfstring{$age$}{\textit{age}} and mass-to-light ratio}\label{s.d.ss.physage}

For virialised galaxies, the FP intrinsic scatter \sigintr must arise from
the conversion between the dynamical quantities in the virial theorem\footnote{
Including the structural coefficient} and the
observables constituting the FP. Indeed, dynamical models based on IFS data
form a mass (or virial) plane consistent with $\sigma_\mathrm{intr} = 0$
\citep{cappellari+2013a}. For these models, the key to eliminating \sigintr is
in the \emph{total} mass-to-light ratio, including dark matter ($\approx 20$~per
cent) and IMF variations \citep{cappellari+2013b}.
However, accurate dynamical models require spatially resolved spectroscopy,
which is still too expensive for large cosmological surveys.

Among the
observables we can obtain from photometry or single-fibre spectroscopy, a number
of works argued that stellar-population properties --- and $age$ in particular
--- are good predictors of the FP residuals \Dr \citep{gregg1992,
prugniel+simien1996, forbes+1998, graves+2009a, falcon-barroso+2011,
springob+2012, yoon+park2020}. 
In particular, \citetalias{deugenio+2021} performed a comparative analysis of
various structural and stellar-population observables, finding that $age$ is the
best predictor of \Dr.
Indeed, this prediction is consistent with the fact that --- for sufficiently
high $\langle SNR \rangle$ --- the $age$ hyperplane has lower observed $rms$
than the FP (Fig.s~\ref{f.rh.plane.a}--\subref{f.rh.plane.b}
and~\ref{f.rh.cores.a}) and the \Dr--age correlation disappears
(Fig.~\ref{f.rh.cores.b}).

In principle, replacing $age$ with $\Upsilon_\star$
brings the hyperplane closer to the virial plane
\citepalias[][their~eq.~2]{deugenio+2021} so it should further reduce \sigintr.
But in practice $\Upsilon_\star$ gives larger observed $rms$ than $age$,
pointing to significantly larger measurement uncertainties --- particularly
so, given that the best-fit coefficient of $\Upsilon_\star$ is 30~per\ cent
smaller (in magnitude) than the coefficient of $age$ (Fig.~\ref{f.rh.fits.c}).
Indeed, we find a median uncertainty on $\Upsilon_\star$ of 0.05~dex for the
1~h mock (and 0.04~dex for the 2~h mock; the corresponding values for $age$ are
just 10~per\ cent smaller). For $\Upsilon_\star$, these estimates are in agreement
with the lowest estimates from \citet{gallazzi+bell2009} and might
indeed be too optimistic. If the larger $rms$ of the $\Upsilon_\star$
hyperplane was due entirely to underestimated measurement uncertainites on
$\Upsilon_\star$, then their median value must be \textit{at least} $0.08$~dex
\citep[more in line with the upper estimate from][]{gallazzi+bell2009}.
Intriguingly, these larger uncertainties are not captured by our method
(which compares repeat observations, using the procedure described in
\S~\ref{s.das.ss.kin.ext} for \sigap). The $\Upsilon_\star$ hyperplane could
represent an independent testing bench to compare the precision of different
spectro-photometric fitting algorithms and modelling assumptions.

Alternatively, $age$ might contain additional information that is not captured by $\Upsilon_\star$. For example, in addition to $\Upsilon$ variations at fixed IMF,
$age$ might correlate with the true stellar $\Upsilon$, i.e., including systematic IMF variations; this could be via the empirical correlations between $age$ and $\sigma$ \citep[e.g.,][]{gallazzi+2005,mcdermid+2015,barone+2018}, and between $\sigma$ and IMF parameters \citep{cappellari+2013b}.
Alternatively, $age$ may contain some structural and/or dynamical information, as evidenced by the fact that $age$ is the best predictor of the spin parameter $\lambda_R$ \citep{croom+2024}, which captures the degree of rotation support in galaxies \citep{emsellem+2011}; after all, early-type galaxies with different values of $\lambda_R$ show different behaviours on the FP \citep{bernardi+2020}.

\subsection{Physical meaning: spectral index}\label{s.d.ss.physind}

Moving to the \spindex hyperplane, we find it to have the same or even smaller
$rms$ than both the FP and even the $age$ hyperplane (Fig.~\ref{f.rh.cores.a}).
However, unlike for the $age$ hyperplane, \Dr still anticorrelates with $age$
(panel~\subref{f.rh.cores.b}). This means that to achieve the same reduction in
$rms$ as the $age$ hyperplane, \spindex draws in part from information that is
not contained in $age$. This raises two pressing questions. First, what is this
information? And second, can we combine it with $age$ to reduce the $rms$ even
further?

After $age$, the second best predictor of the FP \Dr is \afe, the abundance of
\textalpha elements relative to iron \citepalias{deugenio+2021}. This information
could be ostensibly present in \spindex via the $\mathrm{Mgb}$ index.
The feasibility of leveraging \afe to reduce the FP $rms$ has already been
proven \citep{gargiulo+2009}. In \citetalias{deugenio+2021} we argued that
the \Dr--\afe anticorrelation was also present at fixed $age$, which here we
confirm and quantify using PCCs: regardless of how we measure \afe, the
\Dr--\afe anticorrelation exists independently from the \Dr--$age$
anticorrelation, with a correlation coefficient that is roughly one third
the coefficent of \Dr--$age$ (Appendix~\ref{app.afe}).
If \afe was indeed the information \spindex uses, the partial independence of
the \Dr--\afe and \Dr--$age$ relations makes it worthwhile attempting to
combine $age$ and \afe to further reduce the $rms$ compared to the $age$ and
\spindex hyperplanes. Indeed, studying the residuals of the $age$ hyperplane, we
find an anticorrelation with \afe, with $\rho = -0.20$ and $P=1.8 \times 10^{-7}$.
These values place \afe among the best predictors of the residuals of the $age$
hyperplane --- and in first place if we exclude observables that appear (directly
or indirectly) in the hyperplane (e.g., luminosity, surface brightness, size).

Physically, \afe is not a strong driver of
$\Upsilon_\star$, so it is unclear how it can predict \Dr even after controlling
for $age$. Repeating the arguments of \citetalias{deugenio+2021}, \afe may be
capturing age information that is too degenerate for $age$ itself, or,
alternatively, might correlate with some of the other observables that relate
the FP to the virial plane: dark-matter fraction and IMF shape. We note here
the latter would be more natural. This is because, compared to low-dispersion
galaxies, high-dispersion galaxies have both higher \afe
\citetext{\citealp{trager+2000, thomas+2010, mcdermid+2015,
scott+2017, watson+2021, watson+2022} --- but see \citealp{liu2020} for a
different view} and more bottom-heavy IMFs \citep{cappellari+2013b}. The latter
increases the mass-to-light ratio above the average value, thus causing the
\Dr-\afe anticorrelation. In contrast, the sign of this
anticorrelation would require that high-\afe galaxies have the largest
central dark-matter fractions, which runs contrary to the results of dynamical
models \citep{cappellari+2013a}.
Investigating the joint use of \afe and $age$ (or \spindex) is beyond the scope
of this article.

Clearly, other explanations are also possible as to what information \spindex
contains but $age$ does not.

\subsection{Breakdown of the distance uncertainties}\label{s.d.ss.distun}

\begin{figure*}
    \centering
    \includegraphics[trim={0 0 0 1.1cm},clip,width=.9\textwidth]{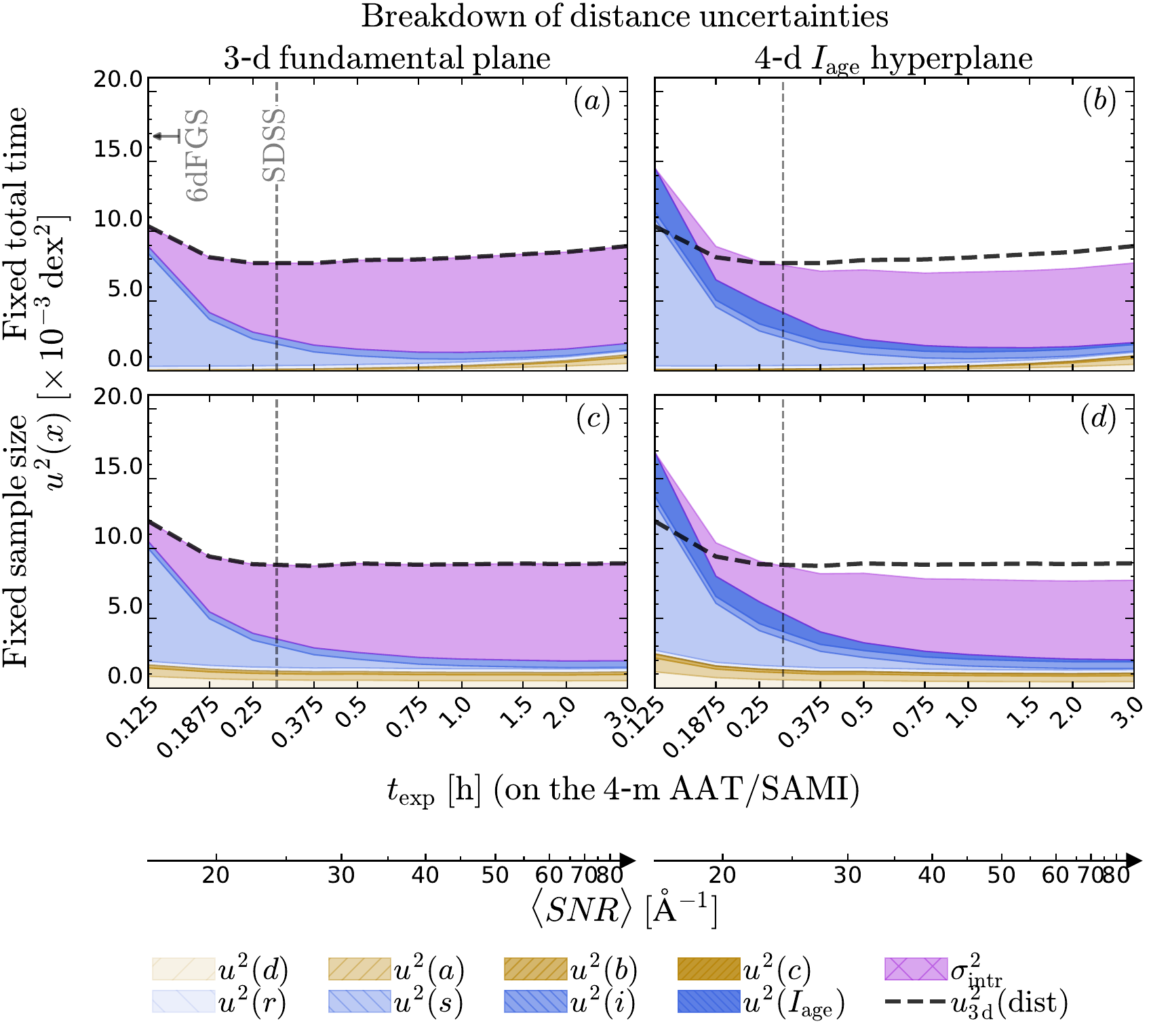}
    {\phantomsubcaption\label{f.d.budget.a}
     \phantomsubcaption\label{f.d.budget.b}
     \phantomsubcaption\label{f.d.budget.c}
     \phantomsubcaption\label{f.d.budget.d}}
    \caption{Breakdown of the median distance uncertainty as a function of
    integration time \texp (and $\langle SNR \rangle$). The vertical dashed line
    is the equivalent exposure time of SDSS; 6dFGS lies outside the range of the
    figure (grey arrow). We show the median distance uncertainty inferred from
    the 3-d FP (left column) and from the 4-d \spindex hyperplane (right column).
    The top row shows the results for a survey of fixed total time; in this
    case, the sample size is set to 5000 for $\texp=3$~hours, and
    increases with decreasing \texp. The bottom row shows the results for a
    survey of fixed sample size equal to 5000 for all \texp.
    The golden regions (from $u^2(d)$ to $u^2(c)$) represent the
    contribution of the uncertainties about the best-fit parameters; the blue
    regions (from $u^2(r)$ to $u^2(\spindex)$) represent the contribution of the
    measurement uncertainties; the purple region on top represents the contribution
    from \sigintr.
    For both large sample size and
    high measurement precision, \sigintr dominates the uncertainty budget. 
    The 4-d hyperplane trades lower \sigintr for
    higher $u^2(c)$ and, especially, $u^2(\spindex)$, the uncertainty in
    \spindex.
    The total uncertainty as a function of \texp is traced by the upper
    envelope; for
    the 3-d case, the upper envelope is also traced by a thick dashed line,
    which we replot for direct comparison in the 4-d case. In panels
    \subref{f.d.budget.b} and \subref{f.d.budget.d}, whenever the
    4-d uncertainty envelope is lower than the thick dashed line, the 4-d
    approach is better than the 3-d approach.
    }\label{f.d.budget}
\end{figure*}

We have seen that, for an exposure time of 1~hour, the 4-d hyperplanes have less
intrinsic and observed scatter than the 3-d FP. The critical question,
however, is whether this translates into increased precision on the derived
distances. The risk, as shown by \citetalias{magoulas+2012} for the $age$
hyperplane, is that large measurement uncertainties on $age$ may frustrate the
improvement in scatter between the 3-d and 4-d relation. Moreover, in most real
applications, higher $SNR$ comes at the expense of smaller sample size, so it is
critical to find the tradeoff between these two aspects of a survey.
To establish whether,
and under what conditions, the 4-d hyperplane yields more precise distances, we
estimate the uncertainty on hyperplane-derived distances by using the error
propagation formula
\begin{equation}\label{eq.r.breakdu}
\begin{split}
  u_\mathrm{4d}^2(\mathrm{dist}) \equiv & \, \sigma_\mathrm{intr}^2
                                          + u^2(d) + u^2(r) 
                                          + \left(u^2(a) s^2 + a^2 u^2(s)\right) \\
                                      + & \left(u^2(b) i^2 + b^2 u^2(i)\right)
                                          + \left(u^2(c) X^2 + c^2 u^2(X)\right)
\end{split}
\end{equation}
where $u(x)$ is the uncertainty on any given quantity $x$,
$X$ is either $A$, \spindex or $\upsilon_\star$, depending on the hyperplane considered,
and the expression for \utd is readily obtained by setting $c=0$ and $u(c)=0$.
The measurement uncertainties on the hyperplane observables $r, s$ and $i$ have been estimated in
\S~\ref{s.das}, whereas the uncertainties on the hyperplane parameters are obtained
directly from the fit (we checked that these values are consistent with the values
obtained by bootstrapping the sample).
The value of the distance uncertainty varies from galaxy to galaxy, so to
simplify our analysis we consider only the median uncertainty value over the
whole sample. In Fig.~\ref{f.d.budget} we show the total uncertainty (dashed
black line) and each of the individual addends; note that we show the
\textit{squares of the uncertainties}, to preserve the linearity of
Eq.~(\ref{eq.r.breakdu}); the uncertainties themselves are shown in
Fig.~\ref{f.d.compare}.
We further note that the parameter uncertainties $u(a), u(b), u(c)$ and
$u(d)$ decrease approximately as the square root of the sample size. For this
reason, for our sample of only 703 galaxies, these uncertainties have a 
disproportionate contribution compared to what they would have for a typical
cosmology application of the FP, which might utilise several thousand galaxies
\citep[e.g.][]{said+2020}. To simulate their relative importance on a
cosmological survey of 5000~galaxies, we divide each of these values by
$\sqrt{5000/703}$; further, we simulate a survey of finite total integration
time. This implies that the sample size decreases with increasing exposure time.
To include this effect, we further multiply the parameter uncertainties
$u(a)$--$u(d)$ by $\sqrt{\texp}$.

The rescaled FP distance uncertainties as a function of \texp are shown by the
dashed black line in Fig.~\ref{f.d.budget.a}. The coloured regions represent the
breakdown of the total uncertainties into each of the constituent terms, so the
upper coloured
region coincides with the dashed black line. The uncertainty on the plane
parameters increases with increasing \texp, due to the decreasing sample size
(golden regions in panel~\subref{f.d.budget.a}). These uncertainties are
dominated by $u(d)$. The typical measurement
uncertainties instead have the opposite trend (blue regions), and are dominated
by $u(s)$. As a result of
these two opposite trends, the total uncertainty \utd has a minimum, at
$\texp \approx 0.25$~hours. This is slightly shorter than the
equivalent exposure time of SDSS (vertical dashed line).

The inconsistent nature of the magenta cross-hatched region, which
traces the contribution of \sigintr, requires explanation. In principle, this should be
independent of \texp; in practice its variation reflects our inability to reliably estimate
the measurement uncertainties: we determine the total observed $rms$,
so \sigintr is effectively the part of the $rms$ not accounted for by the
nominal measurement uncertainties.

Panel~\subref{f.d.budget.b} is the same as \subref{f.d.budget.a}, but for the
\spindex hyperplane. The total uncertainty \ufd is traced by the upper envelope
of the coloured regions, while the dashed black line is \utd from
panel~\subref{f.d.budget.a}, shown for comparison.
We notice immediately that, for $\texp \gtrsim \texp[SDSS]$, we have
$\ufd < \utd$. To our
knowledge, this is the first time that the inclusion of stellar-population
parameters in the FP has been shown to actually improve it as a distance
indicator.

On the other hand, for $\texp \lesssim 0.3$~hours, $\ufd > \utd$, we recover the results of
\citetalias{magoulas+2012} \citep[for 6dFGS, $\langle SNR \rangle = 13$~\AA$^{-1}$,][
so it lies outside the range of Fig.~\ref{f.d.budget.a}]{campbell+2014}.
Looking at the breakdown of the distance uncertainties, we also confirm the
explanation provided by \citetalias{magoulas+2012} as to why, for short \texp,
the FP is a better distance indicator than the 4-d hyperplane. In this case,
the measurement uncertainties on \spindex are too large to compensate the
reduction in \sigintr. Furthermore, the term due to $u(s)$ is also larger than
for the FP case (cf.\ panel~\subref{f.d.budget.a}). This is because, even though
the measurement uncertainties are the same in the two panels, for the 4-d
hyperplane the coefficient of $u(s)$ in Eq.~(\ref{eq.r.breakdu}) is larger
(1.11 vs 1.25, Fig.~\ref{f.rh.plane}, panels~\subref{f.rh.plane.a}
and~\subref{f.rh.plane.c}). Similar considerations apply to the $age$ and
mass-to-light ratio hyperplanes.

Incidentally, Fig.~\ref{f.d.budget} shows that even for data less deep than
SDSS, the median distance uncertainty is dominated by intrinsic scatter
and measurement uncertainties. For a sample size of 5000 galaxies, the
uncertainties on the FP parameters (golden shaded regions) are already negligible
in the total budget. This means that increasing the sample size does not decrease
the median distance uncertainties.

The last two panels of Fig.~\ref{f.d.budget}, panels~\subref{f.d.budget.c} 
and~\subref{f.d.budget.d}, repeat panels~\subref{f.d.budget.a}
and~\subref{f.d.budget.b} respectively, but for a survey of fixed sample size. Compared
to the survey of fixed total time, we simply removed the scaling by
$\sqrt{\texp}$ of the uncertainty in the distance parameters. The
results are qualitatively very similar to the previous case.

\subsection{Implications for cosmology}\label{s.d.ss.implicosm}

\begin{figure}
    \centering
    \includegraphics[width=\columnwidth]{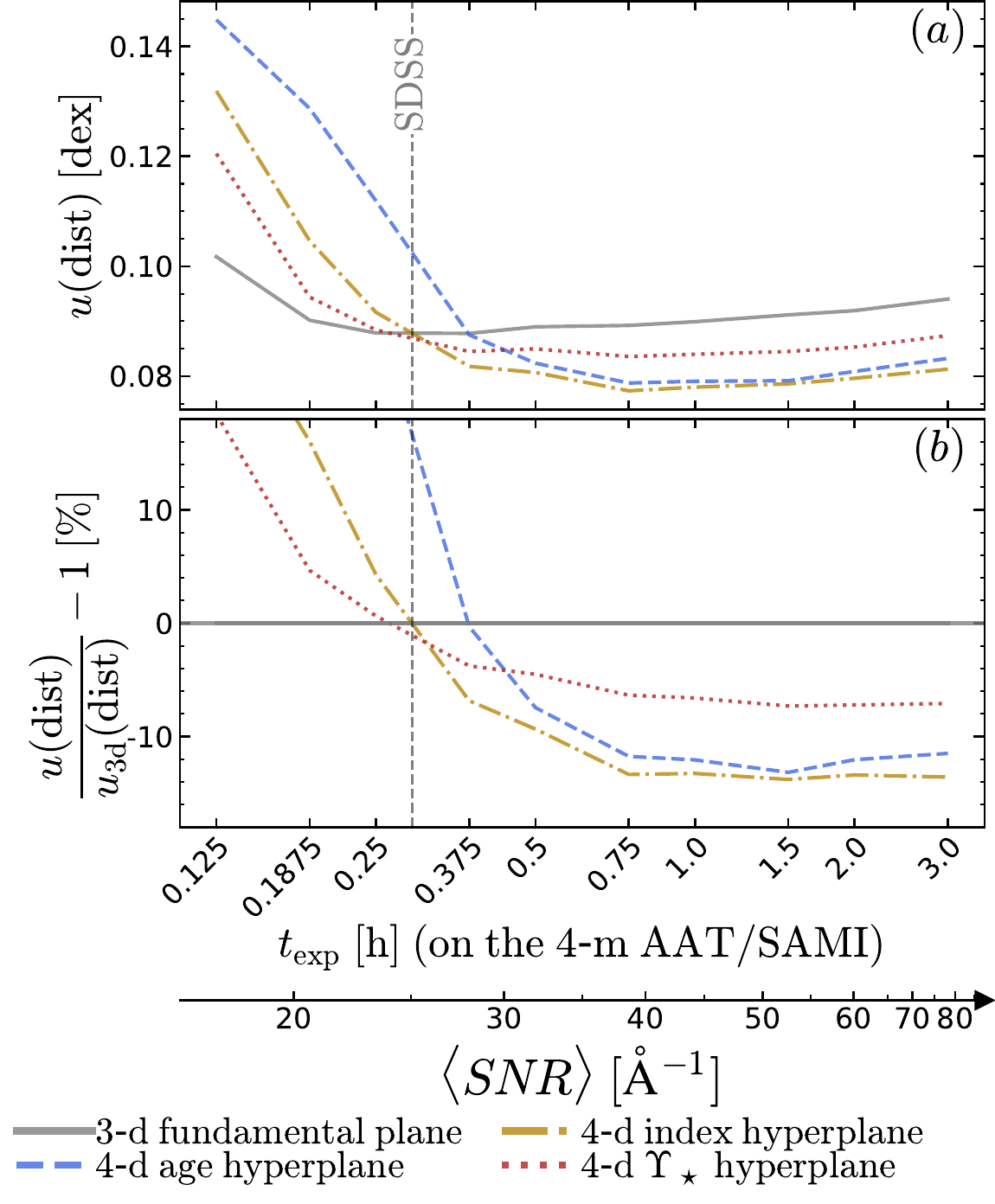}
    {\phantomsubcaption\label{f.d.compare.a}
     \phantomsubcaption\label{f.d.compare.b}}
    \caption{Panel~\subref{f.d.compare.a}: median distance uncertainty as a
    function of exposure time, comparing the FP (solid grey line), to the three
    hyperplanes: the age hyperplane (dashed blue line), the \spindex hyperplane
    (dash-dotted golden line), and the mass-to-light-ratio hyperplane (dotted
    red line). Panel~\subref{f.d.compare.b}: same as
    panel~\subref{f.d.compare.a}, but the median distance uncertainties are
    relative to the value from the FP. For spectroscopy with depth comparable
    to SDSS (vertical dashed line), the \spindex hyperplane is as good a
    distance indicator as the FP; for longer exposure times, the hyperplane
    gives a median uncertainty that is about 10~per\ cent better.
    }\label{f.d.compare}
\end{figure}

In Fig.~\ref{f.d.compare.a} we show again the median distance uncertainty for a
survey of fixed total time, as a function of \texp. The solid grey line is \utd,
the dashed blue line is the \ufd for the $age$ hyperplane, the dash-dotted
golden line is for the \spindex hyperplane, and the dotted red line is for the
mass-to-light ratio hyperplane.
Here we show the uncertainties in their natural units of dex. As expected from
the $rms$ (Fig~\ref{f.rh.cores.a}), for sufficiently long \texp all three
hyperplanes are more precise than the FP. Compared to Fig.~\ref{f.rh.cores.a},
we can see both \utd and \ufd increasing again for the longest \texp, as the
smaller sample size negatively offsets the gain due to lower \sigintr.

Equality between \utd and \ufd is reached first for the $\Upsilon_\star$
hyperplane, followed by the \spindex and $age$ hyperplanes.
By happenstance, equality is reached for \spindex at an
exposure time equivalent to SDSS (vertical dashed line).

The $\Upsilon_\star$ hyperplane has the highest \ufd, despite using
the observable that is most closely related to the physical formulation of the
virial plane \citetext{cf.\ \citetalias{deugenio+2021}, their~eq.~2}, (possibly
due to large measurement uncertainties, \S~\ref{s.d.ss.physage}). Moreover,
$\Upsilon_\star$ shares all the drawbacks of $age$ (e.g., the systematics
inherent to its measurement; see the discussion for $age$ in
\S~\ref{s.das.ss.hfm.ext}), but none of the benefits. Overall, the
$\Upsilon_\star$ hyperplane is the worst of the three candidate replacements for
the FP.

The other two implementations, the $age$ and \spindex hyperplanes, have
substantially lower \ufd, and are comparable to one another. Both of them use a
distance-independent observable as the third independent variable, but \spindex
has the critical advantage of being an empirical observable, measurable directly
from the spectra without any additional assumptions (cf.\ 
\S~\ref{s.das.ss.hfm.ext}).
Even though \spindex varies with distance (at least due to passive evolution),
the same is true for surface brightness; to the extent we use models to correct
the latter, we can apply the same models to correct \spindex too.
These advantages make the \spindex hyperplane a convincing candidate as an
improved distance indicator for cosmology.

Quantitatively, \spindex gives a decrease in the distance uncertainty compared
to \utd of $\approx$0.01~dex. In relative terms, this corresponds to
10~per\ cent of the FP uncertainty, as shown in Fig.~\ref{f.d.compare.b},
where the lines display the relative difference between the distance
uncertainties from Fig.~\ref{f.d.compare.a} and \utd. The maximum improvement
we find is 14~per\ cent.

\begin{figure}
    \centering
    \includegraphics[width=\columnwidth]{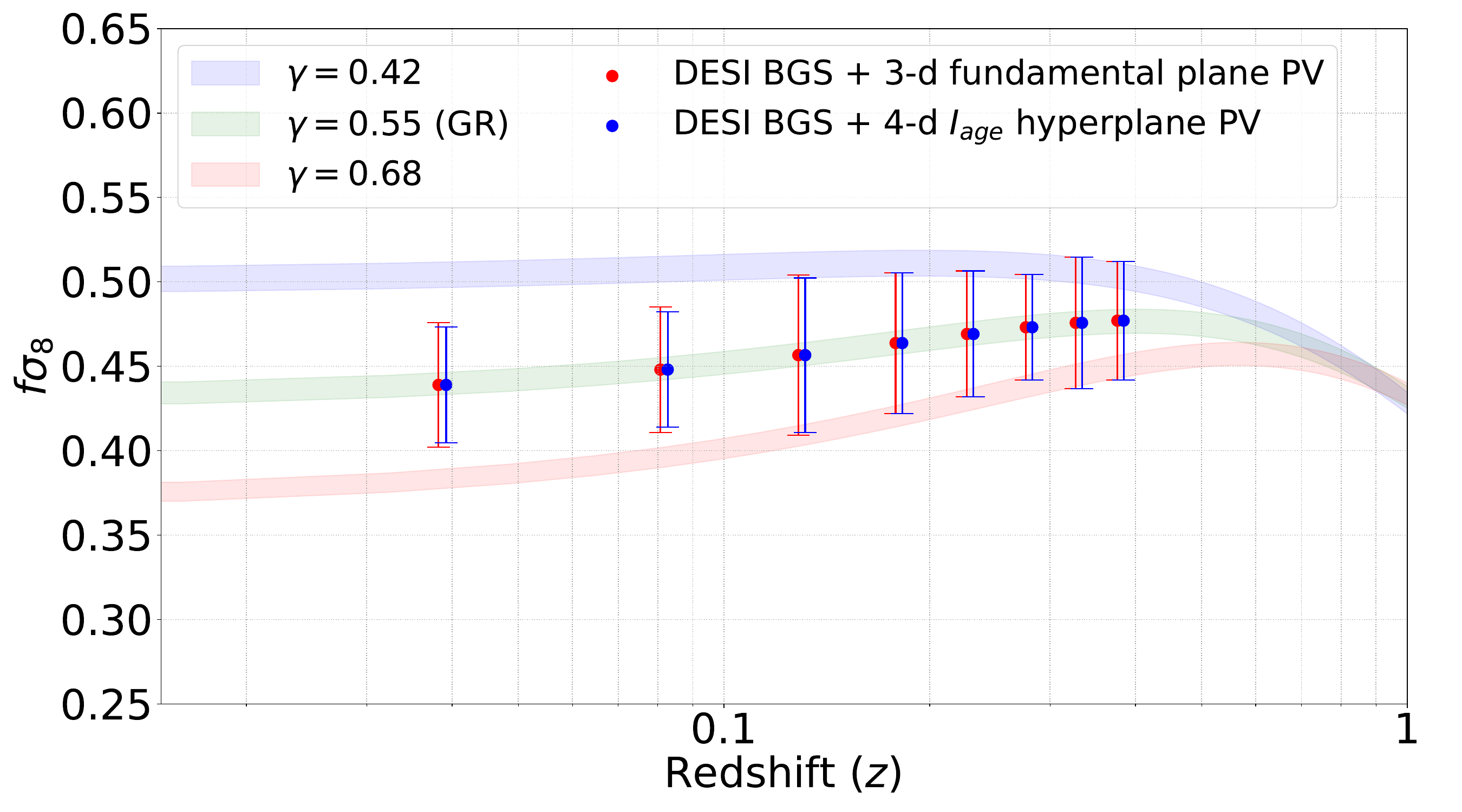}
    \caption{Cosmological Fisher forecasts for the growth rate of structure $f\sigma_8$ from ongoing DESI Bright Galaxy Survey (BGS) and DESI peculiar velocity survey compared to the same survey but using the 4-d hyperplane instead of the 3-d fundamental plane for peculiar velocity. The proposed survey with the \spindex hyperplane would yield 7, 8, and 4~per\ cent smaller uncertainties (blue circles) than the 3-d FP for the first three redshift bins. There is no improvements for the higher redshift bins as most of the constraining power at high redshift are coming from the BGS survey and not from the pv survey. The different bands
    show the expected growth rate of structures from GR (green band) and for
    alternative theories of gravity (blue and red bands).
    }\label{f.d.gr}
\end{figure}

For cosmology, this translates into an equal improvement of the measurement
uncertainties, for example, on the growth rate of cosmic structures
($f \sigma_8$, Fig.~\ref{f.d.gr}). We used the Fisher matrix method of \cite{Howlett2017a} to forecast the constraints on the growth rate of cosmic structure $f\sigma_8$. For this analysis we used the number density from the ongoing Dark Energy Spectroscopic Instrument (DESI; \citealt{Abareshi2022}) as an example. The number densities for the DESI Bright Galaxy Survey (BGS) and for the DESI pv are from \cite{Hahn2022} and Saulder et al. (2022) respectively. Although DESI pv uses both FP and Tully-Fisher for their forecasts, here we use only the number density for the FP galaxies only (see Fig. 11 in Saulder et al. 2022). 

Our Fisher forecasts are plotted in Figure \ref{f.d.gr}. As expected, using the \spindex reduces the uncertainty on $f\sigma_8$ by 7, 8, and 4 per cent for the first three redshift bins, respectively, compared to the FP. The improvement vanishes as we go to higher redshift as most of the constraining power at high redshift is coming from the BGS survey and not from the pv survey. However, for a lower redshift survey such as SDSS pv survey \citep{said+2020} of 8,000 galaxies the reduction of $f\sigma_8$ uncertainly can reach 14 per cent. Compared to SDSS, this proposed survey would
have roughly $1.5\times$ the SNR so $3\times$ longer \texp. In the regime where
sample size is important (i.e., where the golden regions of Fig.~\ref{f.d.budget} are non-negligible), it is better to trade \texp for sample size. However,
when the improvement due to increasing sample size saturates, longer \texp can
still yield a modest improvement to the inferred distances --- and comes with
smaller environment bias too.

However, applying the \spindex hyperplane to SDSS data from \citetext{
\citetalias{howlett+2022}, Fig.~\ref{f.rs.plane.b}}, we find that the residual
correlations between the residuals of the \spindex hyperplane and environment 
are only marginally smaller than the residuals of the FP
(Fig.~\ref{f.rs.plane}, panels~\subref{f.rs.plane.c} and~\subref{f.rs.plane.d}).
This reason for the different behaviour between the \spindex hyperplanes of
SAMI and SDSS is unclear, though we stress that the two dataset have different
measurements of environment (\sigenv for SAMI, \ngroup for SDSS). So it appears
that --- with the current depth of SDSS --- there is little improvement to be
gained from replacing the FP with the \spindex hyperplane.

However, the residuals of the $age$ hyperplane do not correlate with environment,
even when using the SDSS sample and \ngroup (Fig.~\ref{f.rs.plane.f}). Thus
increasing \texp enough to overcome the measurement uncertainties on $age$
provides a distance indicator that is both more precise and accurate than the
FP. This is possible already with the proposed survey with $\texp \approx
3 \times \texp[SDSS]$.

\section{Summary and conclusions}\label{s.c}

In this work, we derive a 4-d scaling relation for early-type galaxies, by
adding an age-related stellar population observable to the 3-d Fundamental
Plane (FP). We considered three of these observables: light-weighted stellar
population age $age$, $r-$band mass-to-light ratio $\Upsilon_\star$, and a new
empirical index \spindex, obtained as a linear combination of Lick indices
(\S~\ref{s.das.ss.hfm.ext}). Crucially, these observables are independent of
distance, so the new relations can be used as distance indicators for near-field
cosmology.
We used a sample of 703 early-type galaxies (selected as closely as possible to
the selection criteria from cosmological FP surveys) and utilised the large
collecting area and long integration times of the SAMI Galaxy Survey on the
4-metre AAT Telescope to find the best tradeoff between long exposure time \texp
and large sample size that minimises the median distance uncertainty for the new
relations.
Our conclusions are:
\begin{itemize}
    \item For the FP, the best-fit coefficients depend on the $SNR$ of the data
        (Fig.~\ref{f.rh.fits}); the dependence is strong (20~per\ cent) for short
        \texp, but disappears for $\texp > \texp[SDSS]$ (here
        $\texp[SDSS] = 0.3$~hours is the exposure time equivalent to that of
        the SDSS survey). For the hyperplanes, the dependency on $SNR$
        vanishes only for $\texp > 2\times \texp[SDSS]$.
    \item The $age$ hyperplane has smaller $rms$ than the FP for
        $\texp > 0.5$~hours (Fig.~\ref{f.rh.cores.a}); the minimum $rms$
        is 0.078~dex (compared to 0.088~dex for the FP). At the same time,
        unlike the FP residuals, the hyperplane residuals \Dr do not correlate
        with $age$ (Fig.~\ref{f.rh.cores.b}). This is evidence for $age$
        information going to reduce the $rms$ compared to the FP.
    \item The $\Upsilon_\star$ hyperplane has smaller $rms$ than the FP already
        for $\texp > 0.3$~hours (Fig.~\ref{f.rh.cores.a}); the minimum
        $rms$ is 0.082~dex (Fig.~\ref{f.rh.cores.b}), making this the worst of
        the three hyperplanes. This disappointing performance may be due to
        the observational uncertainties on $\Upsilon_\star$ being larger than we
        report (we find 0.05~dex from repeat observations, but require at least
        0.08~dex to explain the $rms$ of this hyperplane).
    \item The \spindex hyperplane has $rms$ equal to the FP for $\texp
        \approx \texp[SDSS]$ --- a prediction we verify using SDSS data
        (\S~\ref{s.rs}). For longer \texp, the hyperplane outperforms
        the FP, reaching a minimum $rms$ of 0.077 (Fig.~\ref{f.rh.cores.a}). For
        this hyperplane \Dr still anticorrelates with $age$ (although the
        correlation is weaker and less significant than for the FP;
        Fig.~\ref{f.rh.cores.b}). This suggests that the \spindex hyperplane uses
        in part information other than $age$ to reduce the $rms$.
    \item For the \spindex hyperplane \Dr shows only a marginal correlation with
        environment \sigenv (2-\textsigma), but unlike for the FP, where this
        correlation is highly significant (Fig.~\ref{f.re.resid}). So the
        hyperplane removes the known environment bias of the FP.
    \item Using SDSS data and measuring environment with \ngroup, we confirm
        that the \spindex hyperplane reduces the correlation between the 
        residuals and environment, though the correlation is still significant
        (Fig.~\ref{f.rs.plane.d}).
        In contrast, the residuals of the $age$ hyperplane do not correlate with
        environment (Fig.~\ref{f.rs.plane.f}).
    \item A partial correlation coefficient analysis shows that, comparing $age$
        and \sigenv (or \ngroup), the residuals of the FP correlate
        \textit{independently} with both $age$ and \sigenv (or \ngroup), but
        the correlation with $age$ has by far the largest correlation
        coefficient and highest statistical significance
        (Fig.~\ref{f.re.parcorr} and \S~\ref{s.rs.ss.sdssage})
        This explains why the
        \spindex and $age$ hyperplanes, in correcting the $age$ bias, reduce
        (\spindex) or remove ($age$) the environment bias of the FP.
    \item When simulating a survey of fixed total time, we find that the median
        uncertainty for distances derived from the \spindex hyperplane
        is lower than for the FP for exposure times $\texp>0.3$~hours.
        This result is confirmed by repeating our experiment with data from
        SDSS (\S~\ref{s.rs}).
    \item Compared to the FP, the median distance uncertainty of the \spindex
        hyperplane is up to $\approx$10~per\ cent smaller for SNRs larger than those of SDSS.
    \item Given that \spindex is a distance- and model-independent quantity, the
        corresponding 4-d hyperplane is a potentially superior substitute for the FP as a
        distance indicator for low-redshift cosmology.
\end{itemize}


\section*{Acknowledgments}

We thank the anonymous referee for their insightful comments which greatly improved this article.
FDE and AvdW acknowledge funding through the H2020 ERC Consolidator Grant 683184. FDE
and RM acknowledge support by the Science and Technology Facilities Council (STFC), by the
ERC through Advanced Grant 695671 ``QUENCH'', and by the UKRI Frontier Research grant
RISEandFALL. FDE also acknowledges invaluable help from Brian Van Den Noortgaete.
RM also acknowledges funding from a research professorship from the
Royal Society.
KS acknowledges support from the Australian Government through the Australian Research
Council’s Laureate Fellowship funding scheme (project FL180100168).

The SAMI Galaxy Survey is based on observations made at the Anglo-Australian
Telescope. The Sydney-AAO Multi-object Integral-field spectrograph (SAMI) was
developed jointly by the University of Sydney and the Australian Astronomical
Observatory, and funded by ARC grants FF0776384 (Bland-Hawthorn) and
LE130100198. The SAMI input catalogue is based on data taken from the Sloan
Digital Sky Survey, the GAMA Survey and the VST ATLAS Survey. The SAMI Galaxy
Survey was funded by the Australian Research Council Centre of Excellence for
All-sky Astrophysics (CAASTRO), through project number CE110001020, and other
participating institutions. The SAMI Galaxy Survey website is
http://sami-survey.org/.\\

Funding for SDSS-III has been provided by the Alfred P. Sloan Foundation, the
Participating Institutions, the National Science Foundation, and the U.S.
Department of Energy Office of Science. The SDSS-III web site is
http://www.sdss3.org/.\\

GAMA is a joint European-Australian project based on a spectroscopic
campaign using the Anglo-Australian Telescope. The GAMA input catalogue is
based on data taken from the Sloan Digital Sky Survey and the UKIRT Infrared
Deep Sky Survey. Complementary imaging of the GAMA regions is being obtained
by a number of independent survey programmes including GALEX MIS, VST KiDS,
VISTA VIKING, WISE, Herschel-ATLAS, GMRT and ASKAP providing UV to radio
coverage. GAMA is funded by the STFC (UK), the ARC (Australia), the AAO, and
the participating institutions. The GAMA website is http://www.gama-survey.org/ .\\

This work is based in part on observations made with ESO Telescopes at the La Silla Paranal Observatory
under programme ID 179.A-2004.\\

This work made extensive use of the freely available
\href{http://www.debian.org}{Debian GNU/Linux} operative system. We used the
\href{http://www.python.org}{Python} programming language
\citep{vanrossum1995}, maintained and distributed by the Python Software
Foundation. We made direct use of Python packages
{\sc \href{https://pypi.org/project/astropy/}{astropy}} \citep{astropyco+2013},
{\sc \href{https://pypi.org/project/corner/}{corner}} \citep{foreman-mackey2016},
{\sc \href{https://pypi.org/project/loess/}{loess}} \citep{cappellari+2013b}.
{\sc \href{https://pypi.org/project/ltsfit/}{ltsfit}} \citep{cappellari+2013a},
{\sc \href{https://pypi.org/project/matplotlib/}{matplotlib}} \citep{hunter2007},
{\sc \href{https://pypi.org/project/mgefit/}{mgefit}} \citep{cappellari2002},
{\sc \href{https://pypi.org/project/numpy/}{numpy}} \citep{harris+2020},
{\sc \href{https://pypi.org/project/pathos/}{pathos}} \citep{mckerns+2011},
{\sc \href{https://pypi.org/project/pingouin/}{pingouin}} \citep{vallat2018},
{\sc \href{https://pypi.org/project/ppxf/}{ppxf}} \citep{cappellari+emsellem2004, cappellari2017, cappellari2022},
and {\sc \href{https://pypi.org/project/scipy/}{scipy}} \citep{jones+2001}.
The 4-d hyperplane fits used a modified version of the proprietary
{\sc ltsfit} algorithm; this modified version is available only
for internal use, but may be obtained through a reasonable request to both the
\sendemail{francesco.deugenio@gmail.com,mcx@ox.ac.uk}{Hyperplanes: software sharing request}{corresponding author},
and the \sendemail{mcx@ox.ac.uk}{Hyperplanes: software sharing request}{original developer}.
All computationally intensive calculations were performed using
{\sc \href{https://www.docker.com/}{docker}} \citep{merkel2014} and
{\sc \href{https://htcondor.org/}{htcondor}} \citep{thain+2005}.

\section*{Data availability}

The data used in this work is publicly available through the
\href{https://docs.datacentral.org.au/sami}{SAMI Data Release 3} \citep{croom+2021a}.
Ancillary data is from the \href{http://gama-survey.org}{GAMA Data Release 3}
\citep{baldry+2018} and raw data is from
\href{https://classic.sdss.org/dr7/}{SDSS DR7} \citep{abazajian+2009},
\href{https://www.sdss3.org/dr9/}{SDSS DR9} \citep{ahn+2012} and
\href{http://casu.ast.cam.ac.uk/vstsp/imgquery/search}{VST}
\citep{shanks+2013, shanks+2015}.
The SDSS peculiar velocity sample from \citepalias{howlett+2022} is available
through \href{https://zenodo.org/record/6640513}{Zenodo}. Age, mass-to-light
ratio and \spindex measurements are available by 
\sendemail{francesco.deugenio@gmail.com}{Hyperplanes: data request}{
contacting the corresponding author.}


\bibliographystyle{mnras}
\bibliography{astrobib} 


\appendix

\section{The stellar-mass plane and the colour hyperplane}\label{app.mstarplane}

\begin{figure*}
    \centering
    \includegraphics[width=1\linewidth]{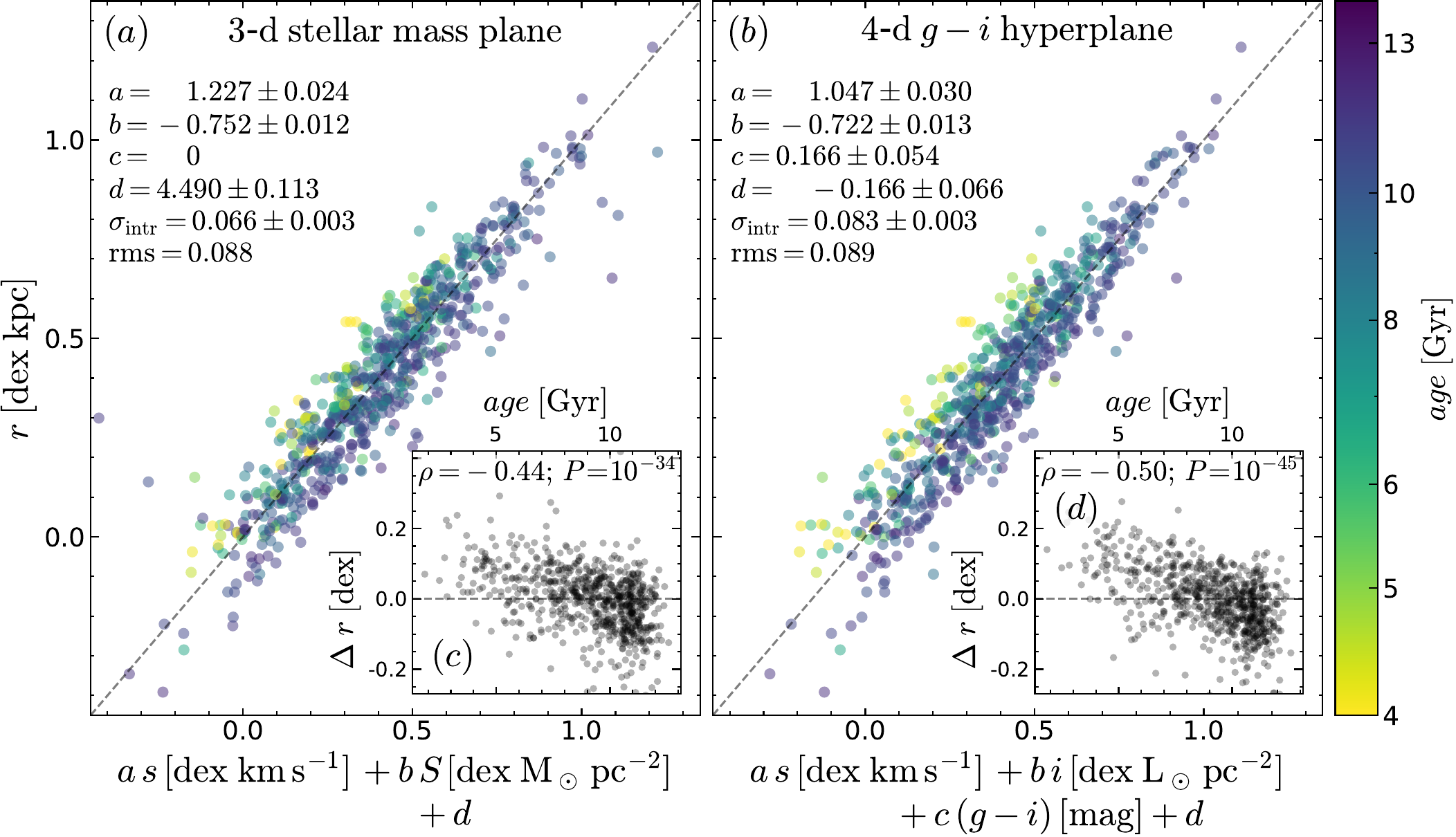}
    {\phantomsubcaption\label{f.a.mstar.a}
     \phantomsubcaption\label{f.a.mstar.b}
     \phantomsubcaption\label{f.a.mstar.c}
     \phantomsubcaption\label{f.a.mstar.d}}
    \caption{Predicted vs measured galaxy size using the stellar-mass
    surface density (panel~\subref{f.a.mstar.a}), and the hyperplane where
    $g-i$ colour is used as fourth variable (panel~\subref{f.a.mstar.b}).
    The stellar-mass plane shows the same $rms$, but lower \sigintr than the FP (Fig.~\ref{f.rh.plane.a}).
    }\label{f.a.mstar}
\end{figure*}

In \S~\ref{s.rh.ss.planes} we briefly argued that the stellar-mass plane is
a viable alternative to the standard FP.
This is shown in Fig.~\ref{f.a.mstar.a}, which has the same y-axis as
Fig.~\ref{f.rh.plane.a}, for ease of comparison.
We use observations with \texp=1~h.
While there is a nominal reduction in the observed scatter
($rms=0.089$ vs 0.088), this is hardly significant; moreover, an accurate comparison
is hard due to the presence of a substantial fraction of outliers in the mas plane
(these outliers are clipped by the \textsc{ltsfit} algorithm, and do not contribute to
the $rms$).
On the other hand, given that $\Sigma_\star$ has larger observational uncertainties than
$i$ \citep{taylor+2011}, the mass plane ends up with a substantial reduction in intrinsic scatter
(from \sigintr=0.082 to 0.075), similar to  $\Upsilon_\star$ hyperplane (0.073;
Fig.~\ref{f.rh.plane.d}). This suggests a more fundamental nature of the stellar-mass plane compared to the FP, in agreement with theoretical results \citep{degraaff+2023}.
We note, however, that the stellar-mass plane's \sigintr is larger than either the \spindex or $age$ hyperplanes values (0.065 and~0.064, panels~\subref{f.rh.plane.b} and~\subref{f.rh.plane.c}).
Ultimately, what matters for the precision of a scaling relation as a distance estimator is the relation's $rms$.
In this case, the stellar-mass plane is also worse than the \spindex and $age$ hyperplanes.
This suggests that approaches using $M_\star$ may be less precise distance indicators than the \spindex or $age$ hyperplanes, at least in the SNR regime considered here.

We remark here that our $M_\star$ measurements are a linear combination of $i$-band magnitudes and $g-i$ colour \citep{taylor+2011,bryant+2015}.
To test if the effect of $g-i$ colour is sufficient or not to reduce the FP scatter, we use $g-i$ as fourth variable in the hyperplane (Fig.~\ref{f.a.mstar.b}).
This approach gives the same $rms$ as the mass plane, and larger intrinsic scatter (\sigintr=0.083), comparable to the FP (\sigintr=0.082; Fig~\ref{f.rh.plane.a}).
This result suggest that the observed reduction in scatter between our FP and mass plane is driven primarily by the use of $i$-band magnitude, rather than colour.
Indeed, both the mass plane and the $g-i$-colour hyperplane display strong and significant trends between the residuals $age$ (panels~\subref{f.a.mstar.c} and~\subref{f.a.mstar.d}).
This suggests that the stellar-mass plane may not be able to remove the environment bias that comes with the FP (\S~\ref{s.re.ss.envage}).
Because SAMI spectroscopy is limited to $g$ and $r$-band, we cannot test if using $i$-band mass-to-light ratios yields a tighter mass plane or $\Upsilon_\star$ hyperplane; this test would require very high SNR spectroscopy in $i$ band, such as is provided by MaNGA \citep{bundy+2015}.
However, this test is beyond the scope of this article.

Recently, there have been efforts to use the mass plane for cosmology applications \citep{dogruel+2023,dogruel+2024}; these authors model simultaneously $s$, $i$, $M_\star$ and S\'ersic index $n$, finding a reduced intrinsic scatter.
In particular, including $n$ in their fit is key to reducing the scatter, and possibly plays a similar role as \spindex or $age$ in our analysis.
However, this reduced scatter does not translate into an equal reduction in the noise on the inferred galaxy distances.
Comparing the mass plane to the \spindex or $age$ hyperplanes requires considering the balance between \sigintr and sample size;
in fact, while the mass plane has worse \sigintr, this could be overcome by increasing the sample size, via including galaxies of all morphological types in the mass plane.
Late-type galaxies, however, also tend to increase \sigintr; therefore, a direct test is necessary to assess the balance between \sigintr, sample size, and morphology.

\section{Contribution of \texorpdfstring{[$\boldsymbol{\text{\textalpha}}$/F\lowercase{e}]}{} to the fundamental plane scatter}\label{app.afe}

The abundance of \textalpha elements relative to iron is known to correlate with
the FP residuals \citep{gargiulo+2009}, even though this anti-correlation is
weaker (smaller-magnitude correlation coefficient) and less significant than the 
anti-correlation with $age$ \citepalias{deugenio+2021}. In our framework, a
direct comparison would be unfair, because precise \afe measurements require
higher SNR than $age$ \citep{liu2020}.
In this section, we compare \afe measurements obtained from the SAMI data
(i.e., without degrading the SNR) to $age$ measurements from the SAMI mocks
with $t_\mathrm{exp} = 1$~hour. The goal is to assess, at least qualitatively,
whether $age$ and \afe play independent roles in producing the FP residuals.

We use two alternative \afe values: measurements based on Lick
indices, taken from the SAMI DR3 \citep[described in ][labelled
$\mathrm{[\text{\textalpha}/Fe]}_\mathrm{Lick}$]{scott+2017}, and our own measurements from
full spectral fitting, obtained using {\sc ppxf} \citetext{following the methodology of
\citealp{vazdekis+2015} and \citealp{liu2020}; labelled simply \afe}.

\begin{figure}
    \centering
    \includegraphics[width=\columnwidth]{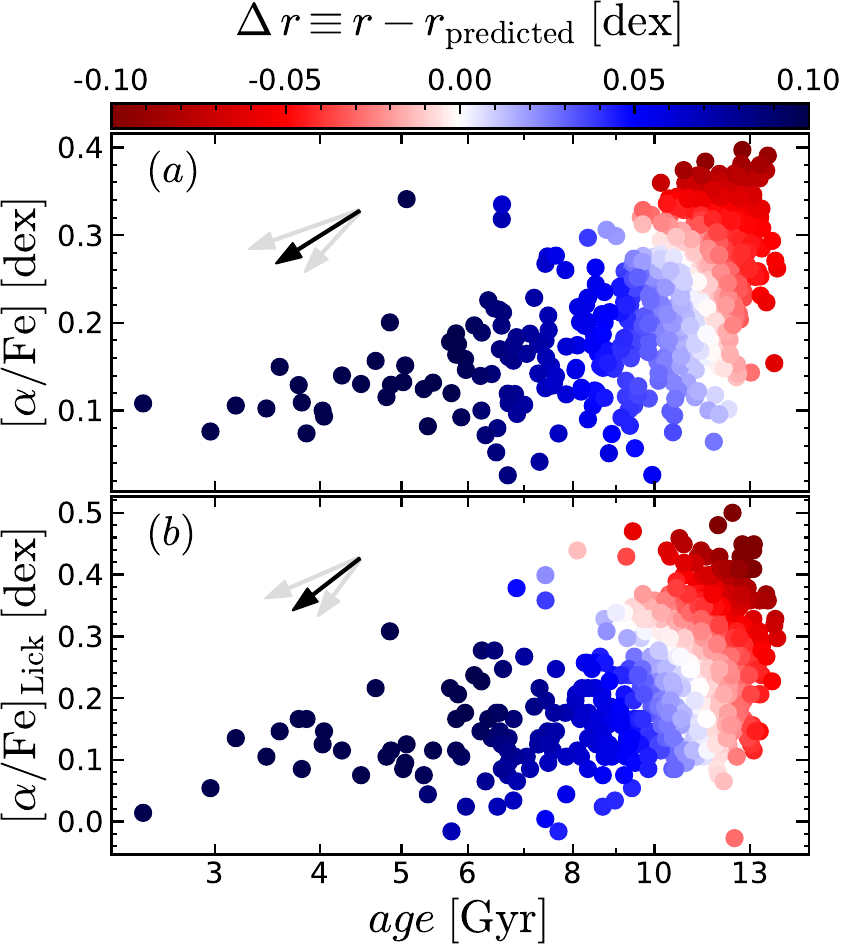}
    {\phantomsubcaption\label{f.app.afe.afe.a}
     \phantomsubcaption\label{f.app.afe.afe.b}}
    \caption{Distribution of the residuals of the 3-d FP as a function
    of both stellar population age and local environment density. The
    residuals have been LOESS smoothed to highlight the trend. The arrow
    is a graphical representation of the partial correlation coefficients;
    it has an angle of $(217\pm14)^\circ$, with the uncertainties
    encompassing the 1\textsuperscript{st} and 99\textsuperscript{th}
    percentiles. An angle of $180^\circ$ would correspond to perfect
    anti-correlation between \Dr and $age$ and no independent correlation
    with \afe. Our findings suggest that $age$ and \afe have independent
    roles in driving the residuals \Dr, with the $age$ correlation
    coefficient being roughly twice the \afe correlation coefficient.
    The two panels show two different estimates of \afe, from full spectral
    fitting (panel~\subref{f.app.afe.afe.a}) and from Lick indices
    (panel~\subref{f.app.afe.afe.b}); both give very similar results.
    }\label{f.app.afe.afe}
\end{figure}

The results are shown in Fig.~\ref{f.app.afe.afe}, where we show \afe
vs $age$, colour-coded by the FP residuals \Dr.
Panels~\subref{f.app.afe.afe.a} and~\subref{f.app.afe.afe.b} show the two
different estimates of \afe, which we find to be in excellent agreement. Regardless
of how \afe is measured, we find that the FP residuals \Dr anti-correlate with both
$age$ and \afe, with older and \textalpha-element enriched galaxies lying
preferentially below the FP. The $age$--\Dr anti-correlation has a larger-magnitude
coefficient and higher statistical significance than the \afe--\Dr
anti-correlation. This is in agreement with \citetalias{deugenio+2021}, but
at variance with \citet{gargiulo+2009}, likely due to their smaller sample size and
lower data quality.
Using partial correlation coefficients, we find that \Dr correlates independently
with both $age$ and \afe, with the $age$--\Dr correlation coefficient being roughly
twice the \afe--\Dr correlation coefficient.
As we already discussed in \citetalias{deugenio+2021}, it is still unclear why
\afe, which is not associated with strong trends in $\Upsilon_\star$, correlates
with the FP residuals. We discuss this subject in \S~\ref{s.d.ss.physind}.


\bsp	
\label{lastpage}
\end{document}